\RequirePackage{fix-cm}
\documentclass[twocolumn]{svjour3}          
\smartqed  
\usepackage{graphicx}
\usepackage[normalem]{ulem}
\usepackage{subfigure}
\usepackage{makecell}
\usepackage{bm}
\usepackage{threeparttable}
\usepackage{amsmath}
\usepackage{multirow}
\usepackage{marvosym}
\usepackage{algorithm}  
\usepackage[noend]{algorithmic} 
\usepackage{xcolor}

\renewcommand{\algorithmicrequire}{\textbf{Input:}}
\begin{document}
\title{DumpyOS: A Data-Adaptive Multi-ary Index for \\Scalable Data Series Similarity Search
}


\author{Zeyu Wang         \and
        Qitong Wang \and
        Peng Wang\textsuperscript{\Letter} \and
        Themis Palpanas \and
        Wei Wang
}

\institute{    Zeyu Wang \at
              School of Computer Science, Fudan University \\
              \email{zeyuwang21@m.fudan.edu.cn}           
           \and
           Qitong Wang \at
              LIPADE, Universit{\'e} Paris Cit{\'e}\\
              \email{qitong.wang@etu.u-paris.fr}
              \and 
               Peng Wang\at
              Shanghai Key Laboratory of Data Science, School of Computer Science, Fudan University \\
              \email{pengwang5@fudan.edu.cn}
              \and
              Themis Palpanas \at
              LIPADE, Universit{\'e} Paris Cit{\'e} \& IUF \\
              \email{themis@mi.parisdescartes.fr}
              \and
              Wei Wang \at
              Shanghai Key Laboratory of Data Science, School of Computer Science, Fudan University \\
              \email{weiwang1@fudan.edu.cn}
}

\date{Received: date / Accepted: date}

\maketitle
\begin{abstract}
Data series indexes are necessary for managing and analyzing the increasing amounts of data series collections that are nowadays available.
These indexes support both exact and approximate similarity search, with approximate search providing high-quality results within milliseconds, which makes it very attractive for certain modern applications.
Reducing the pre-processing (i.e., index building) time and improving the accuracy of search results are two major challenges.
DSTree and the iSAX index family are state-of-the-art solutions for this problem.
However, DSTree suffers from long index building times, 
while iSAX 
suffers from low search accuracy. 
In this paper, we identify two problems of the iSAX index family that adversely affect the overall performance.
First, we observe the presence of a \emph{proximity-compactness trade-off} related to the index structure design (i.e., the node fanout degree), 
significantly limiting the efficiency and accuracy of the resulting index.
Second, a skewed data distribution will negatively affect the performance of iSAX.
To overcome these problems, we propose Dumpy, an index that employs a novel multi-ary data structure with an adaptive node splitting algorithm and an efficient building workflow.
Furthermore, we devise Dumpy-Fuzzy as a variant of Dumpy which further improves search accuracy by proper duplication of series.
To fully leverage the potential of modern hardware including multicore CPUs and Solid State Drives (SSDs), we parallelize Dumpy to DumpyOS with sophisticated indexing and pruning-based querying algorithms.
An optimized approximate search algorithm, DumpyOS-F that prominently improves the search accuracy without violating the index, is also proposed.
Experiments with a variety of large, real datasets demonstrate that the Dumpy 
solutions achieve considerably better efficiency, scalability and search accuracy than its competitors.
DumpyOS further improves on Dumpy, by delivering several times faster index building and querying, and DumpyOS-F improves the search accuracy of Dumpy-Fuzzy without the additional space cost of Dumpy-Fuzzy.
This paper is an extension of the previously published SIGMOD paper~\cite{dumpy}.
\end{abstract}

\section{Introduction}
\label{sec:intro}
Massive data series collections are now being produced by applications across virtually every scientific and social domain~\cite{talking1,talking2,talking3}, making data series one of the most common data types. 
The problems of managing and analyzing large-volume data series have attracted the research interest of the data management community in the past three decades~\cite{DBLP:conf/sofsem/Palpanas16,DBLP:journals/dagstuhl-reports/BagnallCPZ19}. 
In this context, similarity search is an essential primitive operation, lying at the core of several other high-level algorithms, e.g., classification, clustering, motif discovery and outlier detection~\cite{talking2,DBLP:journals/sigmod/PalpanasB19,app1,app2,app3,DBLP:journals/pvldb/0003WNP22}. 

Similarity search aims to find the nearest neighbors in the dataset, given a query series and a distance measure. The naive solution is to sequentially calculate the distances of all series to the query series. 
However, sequential scan quickly becomes intractable as the dataset size increases.
To facilitate similarity search at scale, a data series index can be used to prune irrelevant data and thus, reduce the effort required to answer the queries. 
Moreover, as researchers pay more attention to data exploration, the importance of approximate similarity search grows rapidly~\cite{hydra1,hydra2}. 
It is widely employed in real-world applications since it can provide high-quality approximate query results within the interactive response time, in the order of milliseconds~\cite{nn-workshop,wang-evaluation,approx1,li-evaluation,elpis}. 
In such applications, approximate query result quality is sufficient to support downstream applications~\cite{knn-prediction}.
Recent examples include 
1) a k-nearest-neighbors (kNN) classifier~\cite{knn-classifier}, whose accuracy converges to the best when kNN mean average precision (MAP) reaches 60\%;
2) an outlier detector~\cite{knn-outlier} that achieves the best ROC-AUC with 50\% MAP;
and 3)
a kNN-based SoftMax approximation technique for large-scale classification, which achieves nearly the same accuracy as the exact SoftMax when kNN recall reaches 80\%~\cite{knn-softmax}.
For these applications, the core requirements for the kNN-index are the query time under the above precision (should be in the order of milliseconds), the index building time, and the scalability to support large datasets.

Although there are dozens of approaches in the literature that can index data series~\cite{hydra2},
only a few of them can robustly support large data series collections, e.g., over 100GB (which is why techniques for approximate search~\cite{hydra2}, as well as progressive search for exact~\cite{pros} and approximate~\cite{DBLP:journals/tvcg/JoSF20} query answering have been studied).
Among them, DSTree~\cite{ds-tree} and the iSAX index family~\cite{c19-isip-Palpanas-isaxfamily} show the best query performance on the approximate search and support exact search at the same time.
Due to the dynamic segmentation technique, DSTree requires a long index building time (over one order of magnitude slower than iSAX) and is hard to optimize.  
On the contrary, benefiting from  fast index building and rich optimizations~\cite{isax2+,ads,DBLP:journals/tkde/YagoubiAMP20,paris+,messi,sing,odyssey}, the iSAX index family has become the most popular data series index in the past decade.
Nonetheless, iSAX still suffers from unsatisfactory approximate search accuracy when visiting a small portion of data (one or several nodes, ensuring millisecond-level delay), e.g., its MAP is less than 10\% when visiting one node while query time exceeds one second when improving MAP to $\geq$50\%~\cite{hydra2}.
In this work, we identify the intrinsic problem of the index structure and building workflow of the iSAX index family and propose our novel solution, Dumpy, to tackle those.

First, we observe that the design of the index structure is an inherent but overlooked problem that significantly limits the performance of the iSAX index family.
Although iSAX~\cite{isax} does not in principle limit the fanout of a node, popular iSAX-family indexes~\cite{isax2.0,ads,DBLP:journals/tkde/YagoubiAMP20,ulisse,paris,messi,sing,odyssey} still adopt a binary structure (except for the first layer that has a full fanout).
When a node contains more series than the leaf size threshold $th$, it selects one SAX segment and splits the node into two child nodes.

However, under this binary structure, the splitting policies being used~\cite{isax,isax2.0} often lead to sub-optimal decisions (cf.~\cite{tardis}, Section~\ref{sec:trade-off}), that hurt the proximity (i.e., similarity) of series inside a node, and finally the quality of approximate query results.

Recently, a full-ary SAX-based index has been proposed to tackle this problem~\cite{tardis}.
A full-ary structure splits a full node on all segments, such that it avoids the problems of focusing on a single segment that leads to sub-optimal splitting decisions.
However, it generates too many nodes (at most $2^w$, $w$ is the total number of segments) in each split, leading to an excessive number of leaf nodes, and hence extremely low leaf node fill factors. 
This leads to an underperforming (disk-resident) index, due to inefficient disk utilization and overwhelming disk accesses. 
Although subtrees in the index can be merged into larger partitions (e.g., 128MB)~\cite{tardis} to reduce random I/Os, it still incurs substantial overhead to store and load its large internal index structure and introduce many application limitations at the same time.

We term the aforementioned problems as the \emph{proxi\-mity-compactness trade-off}.
Both proximity and compactness contribute to similarity search since proximity provides closer series to the query and compactness provides more candidate series when visiting a node.

The binary index structure aims at providing compact child nodes, but impairs the accuracy of query results, whereas the full-ary structure splits the node to preserve the proximity of series inside nodes, but fails to provide leaf nodes of high fill factors (i.e., compactness). 
As a result, both structures fail to exploit the proximity-compactness trade-off,
limiting their performance on search accuracy and also building efficiency.

In this work, we break the limits of a single fixed fanout for the iSAX-family indexes and propose an adaptive split strategy that leads to a multi-ary index structure.
Specifically, we design a novel objective function to estimate the qualities of candidate split plans in the aspects of both proximity and compactness.
We use the average variances of data on selected segments to measure the intra-node series proximity and the variance of fill factors of child nodes to measure the compactness.
Moreover, we propose an efficient search algorithm comprised of three speedup techniques to find the optimal split plan according to our quality estimation.

Besides the index structure design, we identify two other problems of the iSAX-index family preventing the best exploitation of the novel adaptive multi-ary index structure.
The first observation is that when the fanout is large (e.g., the first layer in the binary structure and all layers in the full-ary structure), data series are often distributed among the child nodes in a highly imbalanced way, which cannot be entirely avoided, even when we choose the best split plan.
That is, most data series concentrate on only a few nodes while most nodes are slight in size.
It usually leads to a large number of small nodes that impair the performance of the resulting index.
The other problem is that the common iSAX index building workflow splits a node by relying only on the distribution of a tiny portion of data, which actually makes the splitting decisions sub-optimal for the data as a whole. 
For example, iSAX2+ tries to balance two child nodes in splitting according to the first $th+1$ series (i.e., split once it is full), but the final average fill factor is usually less than 20\% as verified in our experiments.

To avoid these two problems, we design a flexible and efficient index-building workflow along with a leaf packing algorithm.
Benefiting from the static segmentation of iSAX, our workflow can collect the global SAX word tables without incurring any additional overhead, and make our adaptive split strategy better fit the whole dataset.
Moreover, our leaf node packing algorithm can pack small sibling leaves without losing the pruning power, contributing to fewer random disk accesses during index building and querying.

In summary, by combining the adaptive split strategy with the new index building workflow, we present our data series indexing solution, Dumpy (named after its short and compact structure).
Dumpy advances the State-Of-The-Art (SOTA) in terms of index building efficiency, approximate search accuracy, and exact search performance, making it a fully-functional and practical solution for extensive data series management and analysis applications.

Moreover, generally as a space-partition-based approach, Dumpy also suffers from a common boundary issue~\cite{nsg,spann}.
That is, the kNN of a query may locate in the adjacent node or subtree and near the partition boundary.
Since we only search one to several nodes, these true neighbors may be missing.
To alleviate this effect, we propose a variant of Dumpy, Dumpy-Fuzzy, which transfers the \emph{hard} partition boundary to a \emph{fuzzy range}, and adopts a duplication strategy in each split to further improve the search accuracy, at the cost of a small overhead on index building and storage. 

Modern parallel hardware, such as multi-core CPUs and NVMe SSDs, becomes pervasive nowadays on commodity machines and data centers. 
To fully exploit the potential of these devices, we extend Dumpy to a well-designed parallel solution, DumpyOS (short for Dumpy On Steroids), to further accelerate the index building and pruning-based querying.
Parallel computing can accelerate near-linearly the adaptive split when indexing, and distance calculation when querying while a proper utilization of SSD can resolve the I/O bottleneck.
Moreover, computations and I/Os are separated in DumpyOS so that they can be designed to overlap with each other leading to even higher performance.

Furthermore, by extending the fuzzy mechanism we design a novel approximate search algorithm DumpyOS-F (short for DumpyOS-Fuzzy), which directly operates on the Dumpy index without any physical duplication of the series in the index.
In contrast to Dumpy-Fuzzy, which statically finds series in the fuzzy boundary for a given node during indexing, DumpyOS-F only works during querying by dynamically finding the series that are close to the query series, but located in different leaf nodes.
Therefore, DumpyOS-F is more flexible to select candidate series with better proximity, and gets rid of the compactness constraint of Dumpy's structure.
As a result, DumpyOS-F achieves better accuracy and avoids the data duplication strategy that Dumpy-Fuzzy uses (and which violates the integrity of the Dumpy index structure).

An extensive experimental evaluation with several synthetic and real datasets shows that DumpyOS (with DumpyOS-F) outperforms the state-of-the-art competitors across all measures.

Our contributions\footnote{A preliminary version of this paper has appeared elsewhere~\cite{dumpy}.} can be summarized as follows.

\noindent(1)
We identify the inherent proximity-compactness tr\-ade-off in the structural designs of the current SOTA iSAX-index family, and demonstrate that it limits the quality of approximate query results, as well as the index building efficiency. 

\noindent(2)
We present Dumpy, a novel multi-ary data series index that hits the right balance of the proximity-compac\-tness trade-off by adaptively and intelligently determining the splitting strategy on-the-fly. 

\noindent(3)
We design a powerful and efficient index-building workflow for the iSAX-index family with a novel leaf packing algorithm to handle data skewness and achieve robust performance.

\noindent(4)
We devise Dumpy-Fuzzy to further improve search accuracy by proper data duplication.

\noindent(5)
We develop DumpyOS, a parallel, materialized time series index, which fully leverages multi-core CPUs and NVMe SSDs to achieve significantly higher performance in both index building and pruning-based querying.

\noindent(6)
We design DumpyOS-F, a novel approximate search algorithm that uses the existing Dumpy index structure with no additional space costs. 
DumpyOS-F leads to more accurate search results than Dumpy and Dumpy-Fuzzy.

\noindent(7)
Our experimental evaluation with a variety of synthetic and real datasets demonstrates that Dumpy and its variants provide consistently faster index building times (up to 5.3x; 4x on average), and higher approximate query accuracy (up to 130\%; 65\% higher MAP on average) than the SOTA competitors, with query answering times in the order of milliseconds.
DumpyOS further achieves on average 3.7x faster building time and 5.8x faster exact query time than Dumpy on multi-core CPU and SSD, and DumpyOS-F is 18\% more accurate than Dumpy on average in approximate search.

\section{Related Work}
\label{sec:related}

\noindent\textbf{[Data series indexes]} 
Dozens of methods have been proposed to index massive data series collections~\cite{hydra1,hydra2}. 
Among these, the SAX-based indexes~\cite{c19-isip-Palpanas-isaxfamily} have gained popularity and achieved SOTA performance. 
Following the initial iSAX~\cite{isax} index, 
iSAX2.0 and iSAX2+ \cite{isax2.0,isax2+} provide faster index building through novel bulk loading and node splitting strategies, 
ADS~\cite{ads} optimizes the combined index building and query answering time, 
ULISSE~\cite{DBLP:journals/vldb/LinardiP20} supports subsequence similarity search, 
SEAnet~\cite{qt} improves query results quality for high-frequency time series using deep learning embeddings, while 
DPiSAX~\cite{dpisax}, Odyssey~\cite{odyssey}, PARIS~\cite{paris+}, MESSI~\cite{DBLP:journals/vldb/PengFP21}, SING~\cite{sing}, and Hercules~\cite{DBLP:journals/pvldb/EchihabiFZPB22} exploit distribution and modern hardware parallelism. 
These indexes all inherit the original binary structure of iSAX, which limits their intra-node series proximity.

ADS~\cite{ads}, as a \emph{query-adaptive} index, builds and materializes only the leaf nodes visited by the examined queries. 
However, in the case of a huge query workload that visits all leaf nodes of the index, ADS becomes the same as an iSAX index, with the same query answering properties. 
On the contrary, as a \emph{data-adaptive} index, Dumpy adapts its structure based on the data collection rather than the queries. 
Therefore, its performance is independent of workloads.
(We omit ADS in the experiments since it is not superior to iSAX2.0 and DSTree~\cite{hydra1}.)

TARDIS~\cite{tardis} first notices the drawbacks of the binary structure and proposes a full-ary structure along with a size-based partitioning strategy to merge different subtrees to be applied in a distributed cluster.
However, TARDIS is only for analyzing a static dataset and the enormous structure decreases the building and query efficiency.
We implement a stand-alone version of TARDIS 
in our experiments.
Coconut~\cite{coconut,DBLP:journals/vldb/KondylakisDZP19} builds a B+-tree after sorting the dataset using the InvSAX representations and gains remarkable performance improvement from sequential I/Os in bulk loading. 
However, the sequential layout on disk will be destroyed by further insertions, and the scan-based exact-search algorithm requires a complete InvSAX table to be kept in memory and the raw dataset in place.
And it seems no easy way to restore the classical tree-based pruning in Coconut.
Hence, we do not include Coconut in our experiments. 

DSTree~\cite{ds-tree} achieves remarkable search accuracy by adopting a highly adaptive summarization EAPCA and increasing the number of segments on the fly.
While the side effect is that DSTree cannot skip costly split operations on raw data series.
Bulk loading algorithms and many other optimizations we mentioned are therefore hard to be applied on DSTree.
As evaluated in our experiments, Dumpy provides higher-quality query results than DSTree even on the static summarization iSAX, with a much faster building time.

\noindent\textbf{[Parallel disk-based indexes]}
Many methods are designed to index and query data series in a parallel environment.
DPiSAX~\cite{dpisax} and TARDIS~\cite{tardis} explore the proper way of data distribution on a cluster of machines.
MESSI~\cite{DBLP:journals/vldb/PengFP21}, SING~\cite{sing} and Odyssey~\cite{odyssey} build and query the in-memory iSAX index in parallel, but they cannot be extended to the disk index. 
PARIS~\cite{paris+} is a disk-resident solution that parallelizes the ADS+ index, which only materializes the SAX words when building the index.
PARIS relies on serial scanning of the raw dataset to support exact search, which leads to an additional cost during query answering time (just like ADS+).

On the other hand, dozens of classical indexes (for very low-dimensional data, i.e., not suitable for data series) have been extended and optimized to a parallel environment with multi-cores and SSDs.
For example, PA-Tree~\cite{pa-tree} optimizes the execution paradigm of B+-Tree on NVMe, TreeLine~\cite{treeline} is designed as an update-in-place key-value store on SSD, and FOR-Tree~\cite{for-tree} optimizes R-Tree on SSD by reducing random writes.

Recently, Zheng et al. proposed DecLog~\cite{declog}, which is a novel decentralized logging technique for time series database management systems.
Declog is also designed based on the merits of NVM to improve the I/O throughput.
However, the problem of data series similarity search is not considered.

\noindent\textbf{[High-dimensional vector indexes]}
According to recent studies~\cite{hydra1,hydra2,DBLP:conf/edbt/EchihabiZP21}, similarity search algorithms for data series and high-dimensional vectors could be employed interchangeably.
Representative algorithms for high-dimensional vector search include proximity graph-based methods~\cite{hnsw,wang-evaluation}, showing excellent query performance on small datasets, but consuming excessive time and memory to build and store the graph.
Now they are not easy to scale in billion-scale datasets in commodity machines~\cite{diskann,nsg}.
Product quantization family methods~\cite{pq,opt-pq,new-pq} achieve better query performance on minute-level near-exact search than data series indexes in advanced research.
However, the building time is still over one magnitude slower than DSTree\-~\cite{hydra2}.
Locality Sensitive Hashing (LSH) family methods~\cite{parsketch,DBLP:journals/kais/LevchenkoKYAMPS21,detlsh}, provide probabilistic guarantees for approximate similarity search, but have been shown to fall behind data series indexes in terms of time and space in the general case~\cite{hydra2}, and even more importantly, cannot support exact query answering.

\section{Background}
\label{sec:background}
We first provide some definitions necessary for the rest of this paper, and then explain the iSAX summarization and index.

\begin{definition}[Data Series]
A data series $s=\{x_1,x_2,\\ \dots,x_l\}$ is a sequence of points where $x_i$ is the $i$-th point and $l$ denotes the length of data series $s$.
\end{definition}

We assume a data series database $db$ contains numerous data series of equal length $n$.
We use the kNN (k-Nearest Neighbor) query to denote a specific similarity search query with an explicit number of nearest neighbors.
\begin{definition}[$k$NN Query]
Given an integer $k$, a query data series $q$ and a distance measure $dist$, a \textbf{kNN Query} retrieves from the database the set of series $R=\{r_1,r_2,\dots,r_k\}$ such that for any series $s \in db \setminus R$ and $r_i \in R$, $dist(r_i,q) \leq dist(s,q)$. 
\end{definition}
The choice of the distance measure depends on the particular application.
Euclidean Distance (ED) is one of the most popular, widely studied and effective similarity measures for large data series collections~\cite{measure}. 
Dynamic Time Warping (DTW)~\cite{dtw} is also widely adopted to analyze data series.
Our solution supports both ED and DTW using the same data structure, like other iSAX indexes.
Besides the exact $k$NN query, the approximate kNN query that accelerates the query processing by checking a small subset of the whole database has attracted intensive interest from researchers. 
The approximate query result, $A=\{a_1,\dots,a_k\}$, is expected to be close to the ground truth result $R$.

\begin{figure}[tb]
\subfigure[SAX of $s$]{
\label{fig:sax3} 
\includegraphics[width=0.47\linewidth]{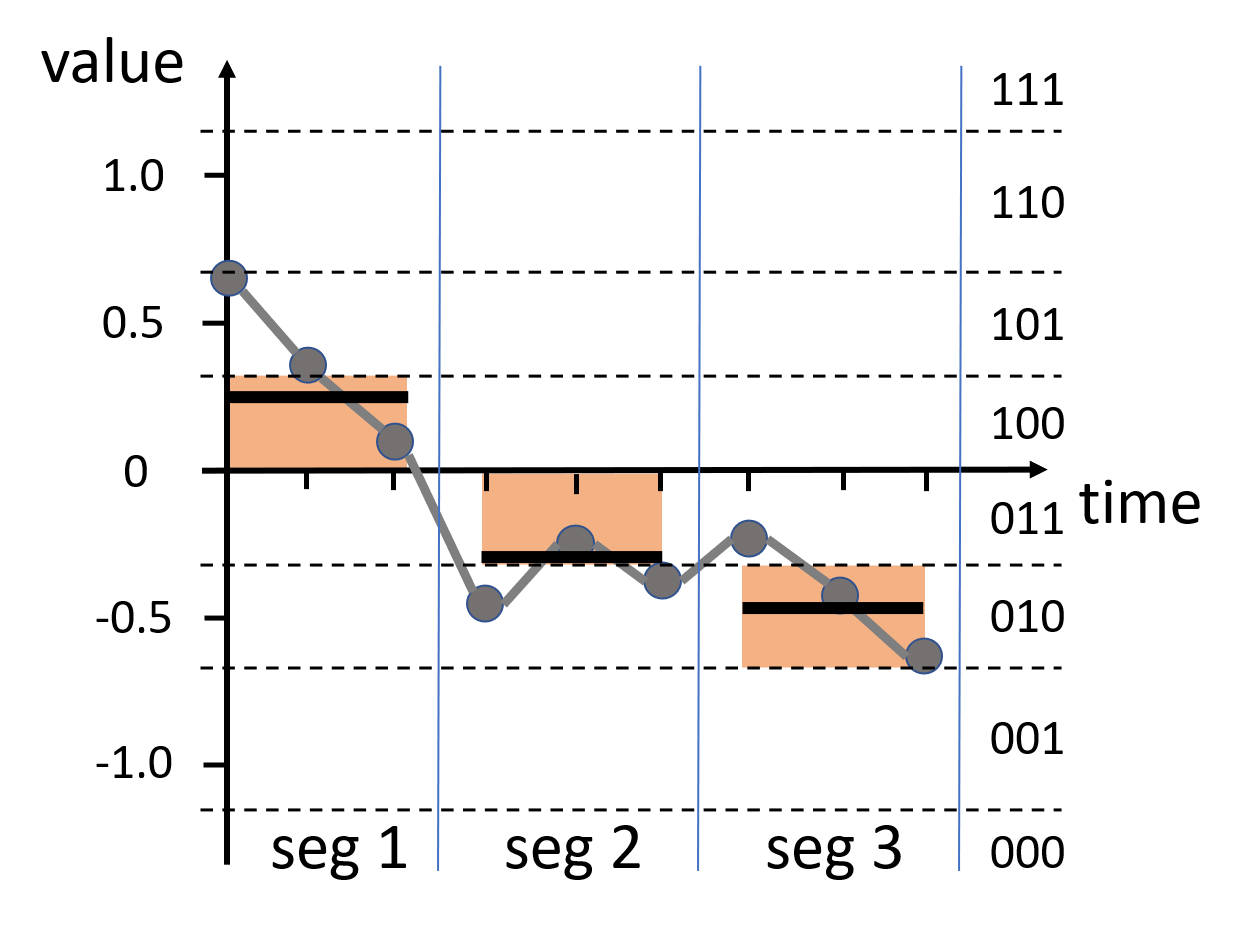}
}\hspace{-2mm}
\subfigure[An iSAX word of $s$]{
\label{fig:sax4} 
\includegraphics[width=0.47\linewidth]{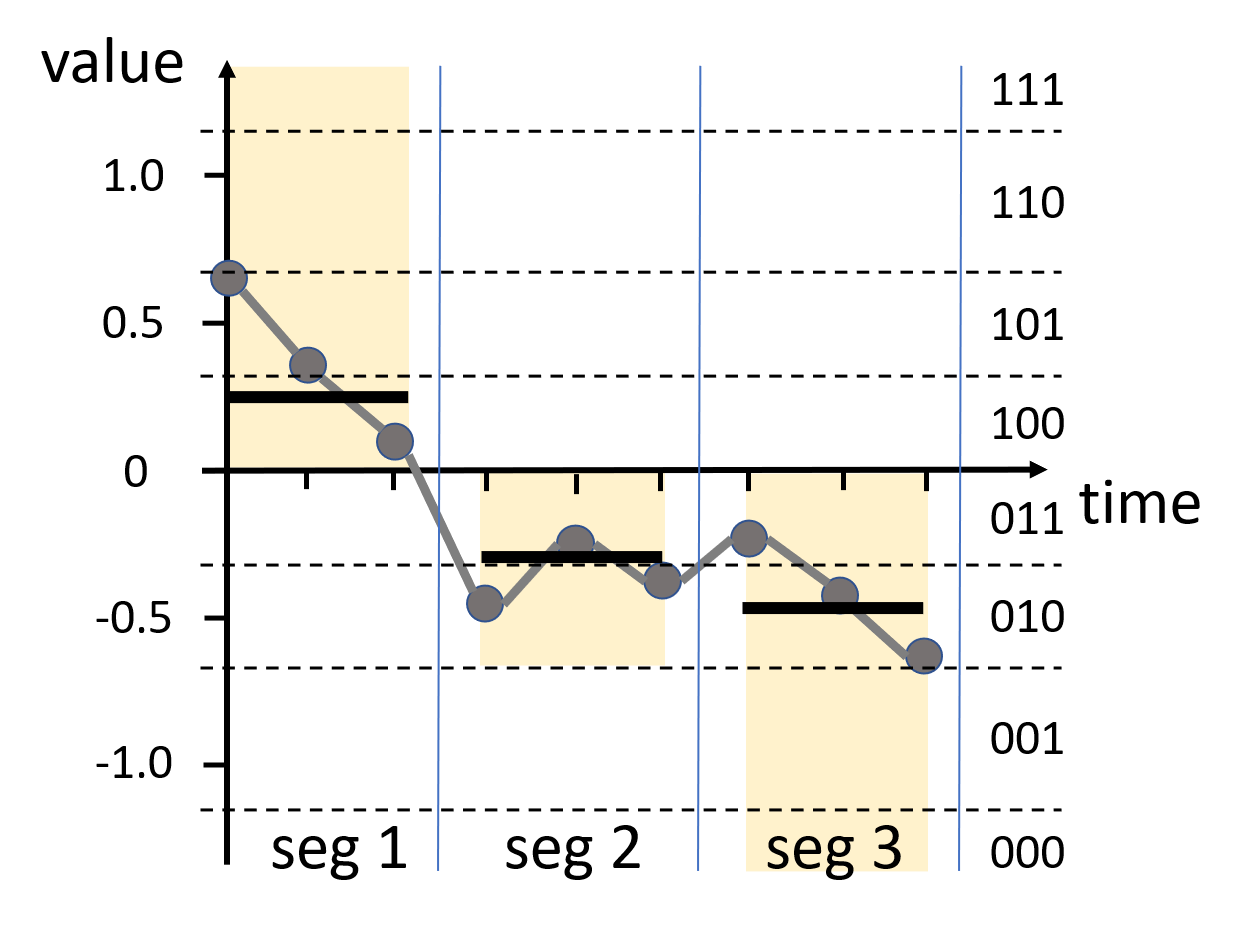}
}

\subfigure[Split on segment 2]{
\label{fig:split2} 
\includegraphics[width=0.36\linewidth]{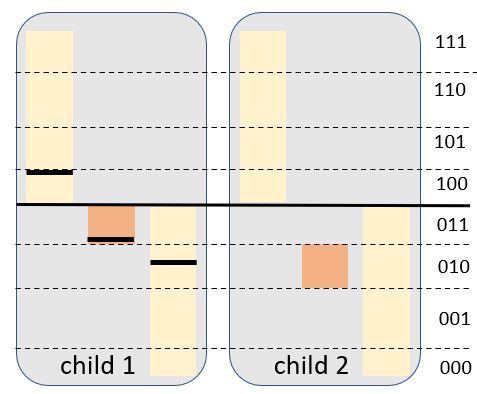}
}
\subfigure[Split on all segments]{
\label{fig:split3} 
\includegraphics[width=0.59\linewidth]{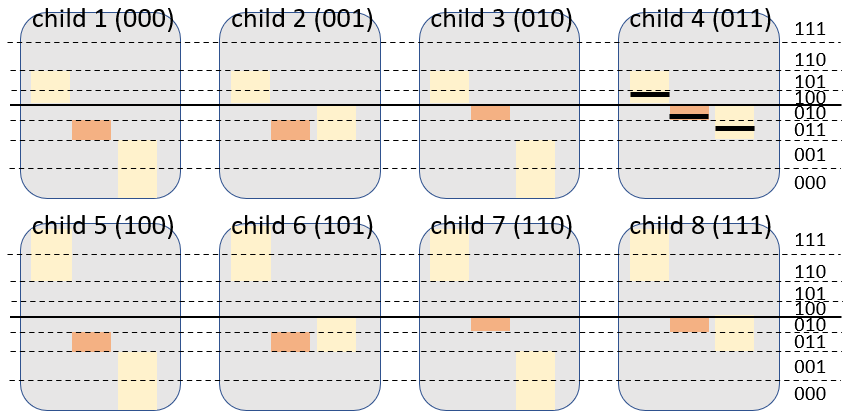}
}
\caption{(a) and (b) are the PAA, SAX and iSAX representation ($w=3$, $b=3$). (c) and (d) are the node splitting for iSAX-index family in two- and full-ary structure, respectively.}
\label{fig:background} 
\end{figure}

\noindent{\bf [iSAX summarization]}
In this paper, we build Dumpy using the iSAX summarization technique~\cite{isax}.
iSAX is a dynamic prefix of SAX words, and SAX is a symbolization of PAA (Piecewise Aggregate Approximation)~\cite{paa}.
We briefly review these techniques with the example in Figure~\ref{fig:background}.

PAA($s$,$w$) divides data series $s$ into $w$ disjoint equal-length segments, and represents each segment with its mean value. 
Hence, PAA reduces $s$ to a lower-dimension\-al summarization.
As the black solid line shown in Figure~\ref{fig:sax3}, PAA($s$,3)=[0.28,-0.31,-0.49].

SAX($s$,$w$,$c$) is the representation of PAA by $w$ discrete symbols, drawn from an alphabet of cardinality $c$.
The main idea of SAX is that the real-value space can be split by $c-1$ breakpoints (subject to $N(0,1)$) into $c$ regions, that are labeled by distinct symbols.
For example, when $c$=4 the available symbols (represented in bit-codes) are \{00,01,10,11\}.
SAX assigns symbols to the PAA coefficients on each segment.
In Figure~\ref{fig:sax3}, SAX($s$,3,8)=[100,011,010]. 
The SAX word represents a \emph{region} formed by the value ranges in $w$ segments, drawn in orange background.
iSAX($s$,$w$,$c$) uses variable cardinality ($\leq c$) in each segment.
That is, an iSAX word is a prefix of the corresponding SAX word. 
The iSAX word in Figure~\ref{fig:sax4} is iSAX($s$,3,8)=[1,01,0]\footnote{A special case for the symbol of iSAX word is $*$, at which segment we use only one symbol $*$ ($c=1$) to represent the whole value range.}.
Due to the decreased cardinality of the alphabet, an iSAX word represents a larger range (more coarse-grained) than the corresponding SAX word.

\noindent{\bf [iSAX index family]}
The iSAX index family~\cite{c19-isip-Palpanas-isaxfamily} uses the tree structure to organize data series, which consists of three types of nodes.
The root node representing the whole value space, points to at most $2^w$ child nodes by splitting on all $w$ segments.
Each internal node contains the common iSAX word of all the series in it, and pointers to its child nodes.
Each leaf node stores the raw data and the complete SAX words of the series inside.
The iSAX index is built by inserting series to the target leaf node one by one.
Once the size of a leaf node exceeds the capacity $th$ (a user-defined parameter), the leaf node gets transferred into an internal node and splits the series inside into several child nodes.
The child node occupies a subspace of the space represented by its parent.
There are two splitting strategies in the iSAX-index family.
The first is the binary split (see Figure~\ref{fig:split2}), which splits a node by doubling the cardinality of the iSAX symbol on \emph{one} segment, and thus, the two child nodes represent disjoint ranges on the specific segment while remaining the same for other segments.

Figures~\ref{fig:sax4} and \ref{fig:split2} show an example.
The node shown in Figure~\ref{fig:sax4} indicates that the series inside the node have a common iSAX word [1,01,0].
If the number of series in this node is more than $th$, it will become an internal node and split the series inside.
With the binary split, it will choose one segment (say, segment 2) and double the cardinality of the iSAX word on this segment.
That is, the iSAX words of the two child nodes are [1,01{\color{green}0},0], [1,01{\color{green}1},0] (see Figure~\ref{fig:split2}).
Then the series inside the parent node is split into these two child nodes according to the value of the second segment.

The second splitting strategy is the full split, where the cardinalities of all the segments are doubled and thus at most $2^w$ child nodes are produced, i.e., [1{\color{green}0},01{\color{green}0},0{\color{green}0}], [1{\color{green}0},01{\color{green}0},0{\color{green}1}] and so on (see Figure~\ref{fig:split3}).
Similarly, the child nodes occupy disjoint subspaces of the parent subspace so that the series inside the parent node will go into the unique child node.
This splitting strategy leads to a completely full-ary structure, adopted by the recently proposed iSAX-family index, TARDIS~\cite{tardis}.

\section{Proximity-Compactness Trade-off}
\label{sec:trade-off}
We now present the proximity-compactness trade-off, based on the analysis of the binary and full-ary index structures.
More specifically, we claim that neither of them can achieve a high leaf node fill factor (i.e., high compactness) and high intra-node series similarity (i.e., high proximity) simultaneously, which limits the index building efficiency and the approximate query accuracy.

\begin{figure}[tb]
\subfigure[Skewed region]{
\label{fig:abnormal} 
\includegraphics[width=0.47\linewidth]{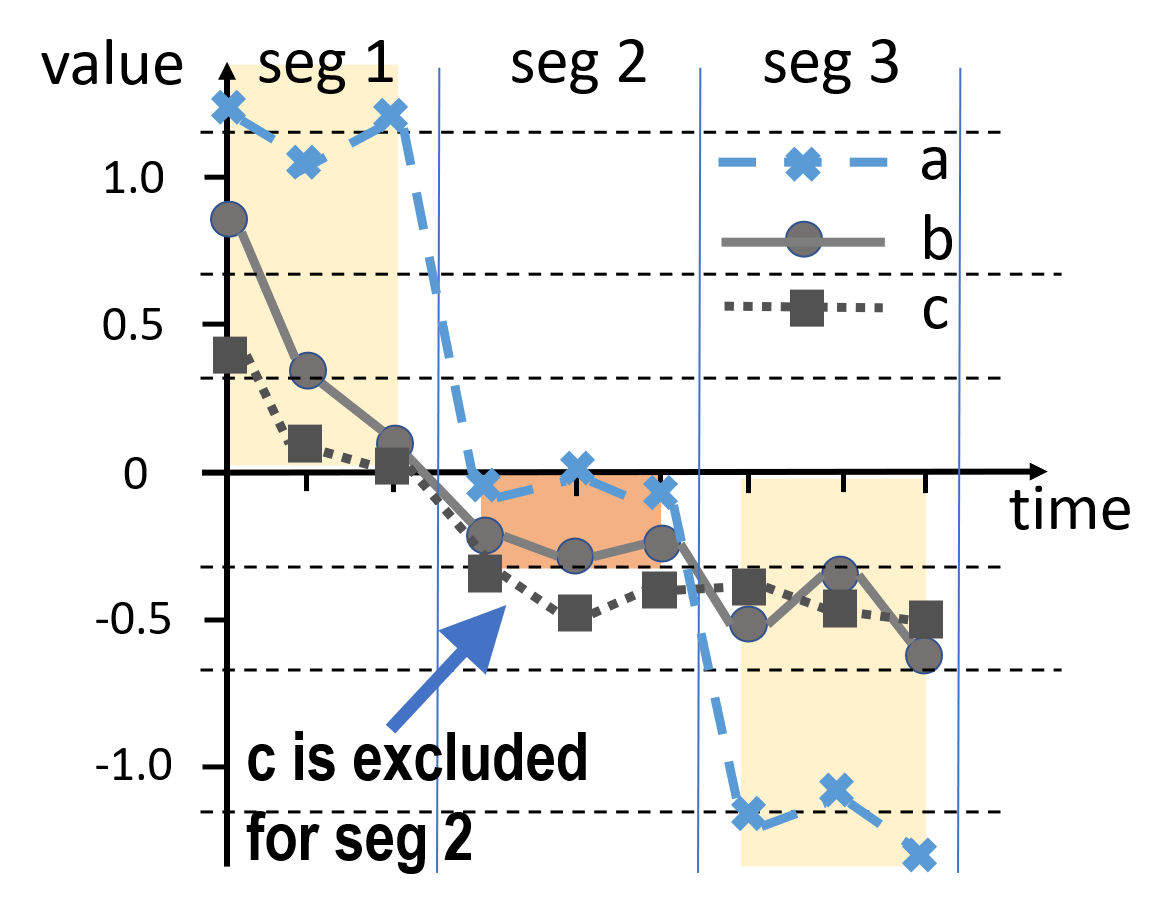}
}
\subfigure[Even region]{
\label{fig:normal}
\includegraphics[width=0.47\linewidth]{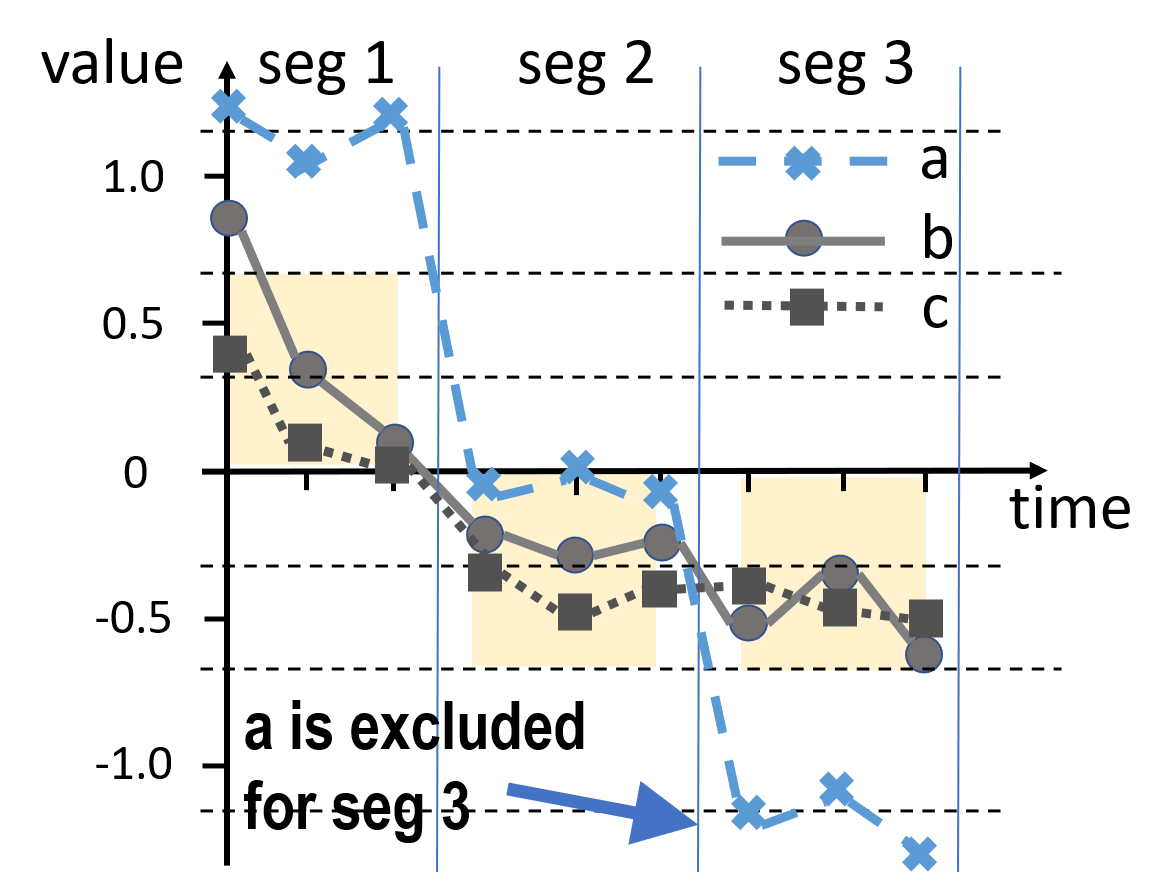}
}
\caption{Illustration of the adverse effect of skewed splits to the intra-node series proximity. Series \emph{b} and \emph{c} are similar to one another, while series \emph{a} is dissimilar to them. In subfigure (a), series \emph{a} and \emph{b} are wrongly grouped in node 1-011-0, whereas in subfigure (b), \emph{b} and \emph{c} are correctly grouped in node 10-01-01.}
\label{fig:curse} 
\end{figure}

\noindent\textbf{[Proximity problem of binary structures]}
In a binary structure index like iSAX, the SOTA splitting strategy~\cite{isax2.0} targets to balance the number of series in the two child nodes, by choosing a segment on which the mean value 
is close to the 
breakpoint.
However, this strategy leads to skewed splits: it may split on several specific segments multiple times, leading to an iSAX word 
with several very high-granularity and other very low-granularity segments. 
This situation is depicted in Figure~\ref{fig:abnormal}, where segment 2 has been split three times, 
while the other segments only once. 
Choosing segment 2 may be the best choice for the parent node, yet, this choice is not beneficial for the overall proximity of the series inside the child node.
In our example, the series \emph{b} and \emph{c} are similar overall, but not grouped together due to the slight difference in segment 2, whereas the distant 
series \emph{a} and \emph{b} are grouped into the same node.

Intuitively, this happens because the split decision considers the similarity of the series in an individual segment (segment 2 in Figure~\ref{fig:abnormal}), while proximity is determined by the overall similarity among series across all segments. 
In other words, all segments should be of approximately the same granularity to better reason about similarity (or equivalently, proximity).
On the contrary, a node with a more even subdivision as in Figure~\ref{fig:normal}, will successfully group series \emph{b} and \emph{c} together.
It is important that, given a binary fanout, no splitting strategy can provide balanced splits while avoiding the skewness problem.
Thus, binary fanout structures inherently suffer from the proximity problem.

\noindent\textbf{[Compactness problem of full-ary structures]}
Contrary to the binary fanout, a full-ary structure~\cite{tardis} splits a node on all segments.
Hence, it intrinsically avoids the skewness problem by creating a strictly even region. 
However, it quickly generates too many small nodes with low fill factors, severely damaging the index compactness. 
Table 1 in our experiments shows the fill factor of a full-ary structure (TARDIS) is below 0.5\% on four public large datasets.
Consequently, the resulting index cannot provide enough candidate series in approximate search, leading to low accuracy when visiting a handful of nodes.
In terms of efficiency, although merging subtrees into larger partitions can significantly reduce random I/Os, storing and loading the enormous structure in a partition file incurs heavy overhead on index building and querying, let alone such dense node packs almost prevent further insertions.

\section{Dumpy}
\label{sec:Dumpy}
In this section, we introduce Dumpy.
Based on a novel adaptive multi-ary structure, Dumpy can hit the right balance of the proximity-compactness trade-off.


\subsection{Index Structure and Design Overview}
\label{sec:motiv}

Dumpy organizes data series hierarchically and adopts top-down inserting and splitting as in other SAX-based indexes.
Once a node $N$ is full (its size $c_N$ exceeds the leaf node capacity $th$), Dumpy adaptively selects $\lambda_N$ segments and splits node $N$ on these segments to generate child nodes.
So the fanout of $N$ $\leq 2^{\lambda_N}$.

We demonstrate an example Dumpy tree with $w=4$ segments in Figure~\ref{fig:pipeline}.
The internal node of Dumpy $N$ maintains a list of chosen segments, $csl(N)=[cs_1,cs_2,\\ \dots,cs_{\lambda_N}]$ where $cs_i$ is the \emph{id} of segments (numbered from 1 to $w$), and $csl(N)$ is sorted by the \emph{id} of segments in ascending order. 
When we concatenate the increased bit of each symbol on $csl(N)$, we can get a $\lambda_N$-length bit-code, denoted by \emph{sid} in Dumpy.

In the physical layout, a leaf node corresponds to continuous disk pages storing the raw series and SAX words.
An internal node maintains a hash table to support tree traversal, named \emph{routing table}, mapping \emph{sid} to its corresponding child node.

We now present the intuitions behind our adaptive node splitting algorithm.
To find the best balance between the proximity-compactness trade-off, we design an objective function to evaluate each possible split plan, where we use the variance of data on certain subspace to estimate the proximity of series inside child nodes and use the variance of fill factors of leaf nodes to surrogate the compactness.
Considering the whole search space is $2^w$+1, we first eliminate unpromising plans and then employ the relationship between different split plans to accelerate searching (cr. Section~\ref{sec:split}).

To better facilitate our adaptive splitting algorithm, we propose a new index-building workflow based on the information of all series (cr. Section~\ref{sec:workflow}).
The building workflow of previous SAX-based indexes 
split a node once it is \emph{just} full, i.e., the $th$+1 series arrives.
Considering a node in the index may be mapped by much more series than $th$, the conventional split decisions will lose effect as the first $th+1$ series soon become a small portion of all series falling into this node.

\begin{figure}[t]
\subfigure[Random walk]{
\label{fig:dist-rand} 
\includegraphics[width=0.47\linewidth]{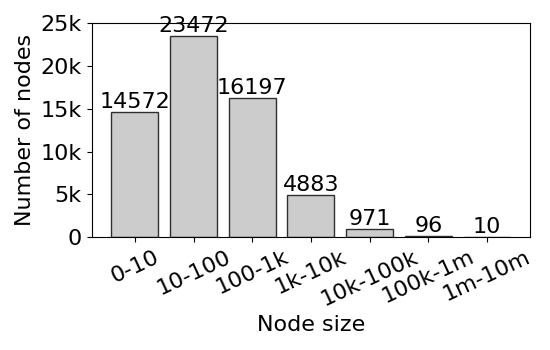}
}
\subfigure[DNA]{
\label{fig:dist-dna} 
\includegraphics[width=0.47\linewidth]{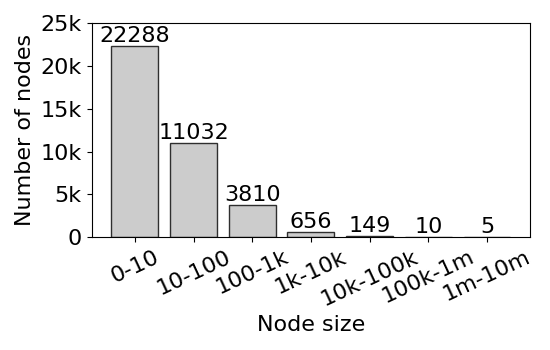}
}
\caption{Node size distribution in the first layer on two 100GB datasets ($w=16$).}
\label{fig:dist-toplayersplits} 
\end{figure}

Last but not least, even if supported by the optimal 
adaptive splits,
there might still exist a large number of small leaf nodes.
This is coming from the fact that data series, similar to high-dimensional vectors, are usually unevenly distributed,
generating many different dense and sparse regions~\cite{DBLP:journals/tkde/KornPF01}.
Figure~\ref{fig:dist-toplayersplits} shows the node size distribution in the first layer of iSAX-type indexes.
$>$60\% nodes in Rand and $>$80\% nodes in DNA have $<$100 series while $<$2\% nodes cover $~$80\% series.
To fully avoid this problem, we propose a novel leaf node packing algorithm, to provide high-quality leaf packs by bounding the maximal demotion bits of them (cr. Section~\ref{sec:pack}).

\subsection{Workflow of Dumpy Building}
\label{sec:workflow}

\begin{figure}[tb]
  \includegraphics[width=1\linewidth]{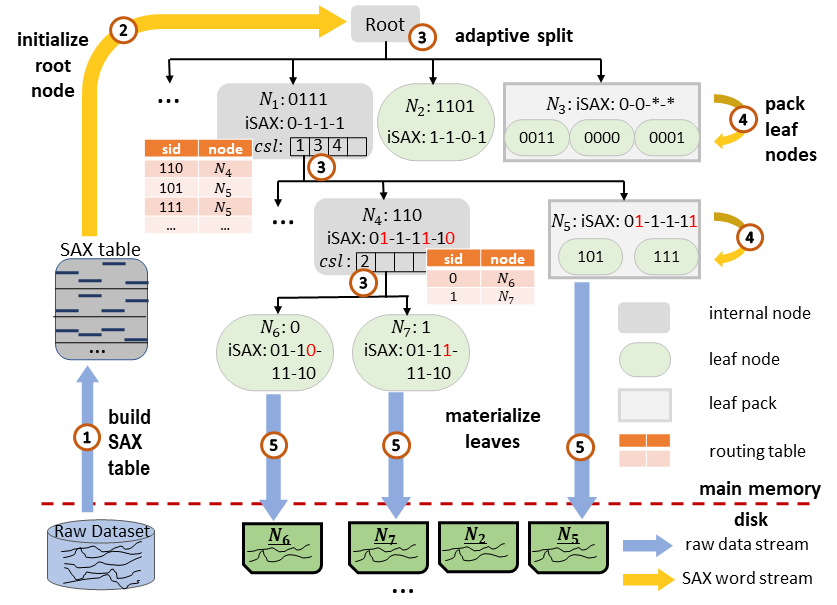}
  \caption{Index structure ($w$=4) and building workflow.
  }
  \label{fig:pipeline}
\end{figure}

The index-building workflow of Dumpy is demonstrated in Figure~\ref{fig:pipeline}.
Dumpy follows the most advanced building framework of the iSAX-family index~\cite{ads-full} but changes two key designs.
The classical framework is a two-pass procedure.
In the first pass, it reads data series in batch from the raw dataset and computes the SAX words of each series.
Then the SAX words are inserted into the destination leaf node one by one and nodes will be split once it is full.
After the first pass, the index structure is in its final form.
In the second pass, data series are again read in batch, routed to the correct leaf nodes, and written to corresponding files finally.

\noindent{\textbf{[Split nodes using a complete SAX word table]}} 
One key point of this framework is to only keep the SAX words in the index (the first pass) and use the SAX words to split nodes, which takes full advantage of the static property of iSAX summarization and significantly reduces disk I/Os.
Dumpy further extends this workflow by separating the SAX words collection and node splitting into two non-overlapping steps, i.e.,
only collecting all SAX words into a SAX table in the first pass and then using
the SAX table to build the index structure before the second pass.
Hence, when splitting a node, 
we know the exact size and the distribution of the series inside it, making our adaptive splitting algorithm take effect actually as we expect.

\noindent{\textbf{[Write to disk after leaf node packing]}}
A large fanout usually generates numerous small nodes before leaf node packing, as shown in Figure~\ref{fig:dist-toplayersplits}.
Considering in the second pass we flush the series of each relevant leaf node in a batch, the number of leaf nodes approximately decides how many random disk writes per batch.
To reduce random writes, before materializing leaves as in the second pass, Dumpy merges sibling small leaf nodes in a proper way to be bigger packs and builds a routing table for the internal node.
Then in the second pass, the series will directly be routed to the leaf pack by the routing table, largely reducing the random disk writes.

\subsection{Adaptive Node Splitting} 
\label{sec:split}
We now present our adaptive strategy of determining fanouts and splits (on which segments) based on the SAX words of all relevant series.
Our strategy is to select the best split plan based on a novel objective function, which considers the proximity of series inside child nodes and the compactness of child nodes at the same time.
Since the number of all possible split plans is as large as $2^w-1$, we also propose an efficient search algorithm by restricting the candidate space and reusing the shared information.



\subsubsection{Objective Function}
Our objective function targets to achieve the best balance between the trade-off of proximity and compactness.
We measure the proximity based on the average variances of data on candidate segments.
To measure the compactness of the children nodes after a split, we consider both the variance of fill factors of child nodes and the ratio of overflowed nodes to pursue a balanced split and avoid bias for small or large fanouts.

Given a node $N$ containing $c_N$ series $\mathcal{X}_{N} = \{\boldsymbol{x^1},\boldsymbol{x^2},\\\dots,\boldsymbol{x^{c_N}}\}$ where $\boldsymbol{x^i}=SAX(s^i,w,c)$ is the SAX word of series $s^i$ and $x^i_j$ is $j$-th symbol of $\boldsymbol{x^i}$, and a split plan $csl(N)=[cs_1,cs_2,\dots,cs_{|csl(N)|}]$, we first project each series of $\mathcal{X}_{N}$ onto the segments of $csl(N)$ and get $\mathcal{X'}_{N}$, that is, $x'^{i}_{j}=x^{i}_{cs_j}$ for any $i$ and $1\leq j\leq|csl(N)|$.
Then our objective function is as follows:
\begin{equation}
\label{equ:score}
\mathop{max}\limits_{csl(N)}\ e^{\sqrt{\frac{1}{|csl(N)|}Var(\mathcal{X'}_{N})}} + \alpha*e^{-(1+o)\sigma_{\boldsymbol{F}}}
\end{equation}
\noindent where $e$ is the Euler's number, the variance of projected data is defined as $Var(\mathcal{X'}_{N})=\frac{1}{c_N}\sum_{i=1}^{c_N}\Vert\boldsymbol{{x'}^i}-\boldsymbol{\mu}\Vert^2$ and $\boldsymbol{\mu}$ is a vector of mean values of data on each chosen segments\footnote{We use the midpoint of the range represented by the SAX symbol to calculate the mean value and other statistics.}, $o\in[0,1]$ is the ratio of overflowed child nodes (size $>th$), $\sigma_{\boldsymbol{F}}$ is the standard deviation of the fill factors of child nodes, and $\alpha$ is a weight factor to balance the influence of these two measurements.

The first term estimates the proximity of a split plan.
It evaluates the average variance of relevant data series on the projected SAX space, which is equivalent to the average distance of all the data series to their centroid, i.e., the mean vector $\boldsymbol{\mu}$.
The variance is an indicator of data informativeness on certain dimension~\cite{opt-pq,new-pq,kd-tree}, considering large variances usually mean large information entropy~\cite{entropy}.
Since different plans may choose different numbers of segments, we divide the variance by the number of chosen segments to make the evaluation fair.

The second term is to evaluate the compactness of a split plan.
The standard deviation of fill factors of child nodes prevents extremely imbalanced splits and avoids the severe data skewness like Figure~\ref{fig:dist-toplayersplits}: the value will be very large in this case.
Informally, the vector of fill factors is defined as $\boldsymbol{F} = (F_1,F_2,\dots,F_{2^{|csl(N)|}})$ where $F_i = c_{N_i} / th$ and $N_i$ is the $i$-th child node. 
However, it shows bias for small fanout, which generates fewer but larger child nodes and leads to an unnecessary deep tree.
To resolve this problem, we add a penalty term (1+$o$) that uses the ratio of overflowed child nodes to avoid the bias for plans of small fanout.


\subsubsection{Find the Optimal Split Plan}
To reduce the complexity of finding the optimal split plan under our objective function, we propose a novel searching algorithm composed of three practical speedup techniques, that are, pre-computing the variance for each segment, restricting the search space by a user-defined fill-factor range, and hierarchically computing the sizes of child nodes.

\noindent{\textbf{[Pre-compute variance]}}
We find that in the first term of the objective function, $Var(\mathcal{X'}_{N})$ can be computed by linearly accumulating the variance of data on each segment.
\begin{equation}
Var(\mathcal{X'}_{N})=\sum_{cs\in csl(N)}Var(\Pi_{cs}(\mathcal{X}_N))
\end{equation}
\noindent where $\Pi_{cs}(\mathcal{X}_N)$ indicates the projection of $\mathcal{X}_N$ onto segment $cs$.


Hence, we can pre-compute the variance of data series on each segment when we start to split a node.
When evaluating a specific plan, we simply fetch the corresponding segments' variances and sum them up with constant complexity.

\noindent{\textbf{[Restrict the search space]}}
The second speedup technique is to restrict the average fill factor of child nodes to be in a reasonable range and avoid the particular evaluation.
We introduce a pair of parameters $F_l$, $F_r$ to bound the average fill factor of child nodes.
Then the range of the number of chosen segments $|csl(N)|$ can be deduced as 
\begin{equation}
\label{equ:fill}
\max(1, \log{\frac{c_N}{F_r*th}}) \leq |csl(N)| \leq \min(w, \log{\frac{c_N}{F_l*th}})
\end{equation}
In practice, we empirically set $F_l=50\%$ and $F_r=300\%$, which generally achieves 16x speedup and 99\% accuracy on average.

\noindent{\textbf{[Hierarchically compute sizes of child nodes]}}
So far for each plan, we still need to iterate all the data series to get the sizes of child nodes.
If a split plan $csl^i(N)$ is a subset of another plan $csl^j(N)$, then the size distribution of child nodes of plan $csl^j(N)$ can be reused to calculate the distribution of plan $csl^i(N)$.
Since the whole $w$ segments are a superset of split plans, 
we first compute child node sizes for $w$ segments as a base distribution in each split
and then traverse other plans in a depth-first manner, starting from the plan with the largest fanout to the smallest.
Hence, we can reuse the size distribution we have gained
in a hierarchical way and avoids 
traversing all the series for each plan.

\subsection{Leaf Node Packing}
\label{sec:pack}
In this subsection, we propose a simple yet effective algorithm to pack small leaf nodes without losing the pruning ability of packed nodes.
The intuition is to minimize the value range in the SAX space occupied by the packed nodes, i.e., make them have the tightest iSAX representation. 
Tighter iSAX representation directly translates to higher pruning power.
We define the demotion bits as the different bits between the $sid$s of two or more nodes considered to be merged into the same pack.
In our node packing algorithm, we limit the number of demotion bits to be smaller than $\rho \lambda$, where $\rho$ is a user-defined parameter trading off pack quality and fill factor.
Specifically, given a list of packs and a small node $N$ to be packed, we decide $N$'s belonging by the demotion cost, which is defined as the increased number of demotion bits of the pack if we add $N$ into it.
A leaf node pack forbids any node's insertion request if it will make the pack demote more than $\rho\lambda$ bits or overflow (size $>th$).
Finally, if  no existing pack can satisfy the requirements to insert $N$,
we will create a new pack and insert $N$ into it.
The details of our leaf packing algorithm can be found in ~\cite{dumpy}.

\subsection{Search Algorithm}
\label{sec:search}
Dumpy supports 
two styles of query-answering algorithms.
The first style follows the classical pruning-based search algorithm~\cite{hydra2}.
As a SAX-based index,
Dumpy can conduct an efficient search (including the exact, $\delta$-$\epsilon$-approximate search and etc.~\cite{isax,hydra2}) by pruning irrelevant leaf nodes using lower-bounding distances of iSAX words~\cite{isax,hydra2}.
Besides that, Dumpy also supports traditional approximate search, i.e., querying one target leaf node.
Moreover, we extend it to allow searching more nodes, called \emph{extended approximate search}, to improve query answer quality while maintaining response time in milliseconds.
We limit the search range of extended approximate search in the smallest subtree of the target leaf node to reduce the complexity and avoid traversing the whole tree and evaluating the nodes one by one as in the bound-based search style.
Benefiting from Dumpy's multi-ary structure and fill factor, it brings prominent improvement in search accuracy.
The details of our algorithm is shown in Algorithm~\ref{alg:search}.


\begin{algorithm}[tb]
\caption{Extended Approximate Search} 
\label{alg:search}
{\footnotesize
\begin{algorithmic}[1]
\REQUIRE root node $N_r$, node number $nbr$, query series $q$
\STATE node $N$ = $N_r$
\WHILE{$N$!= null and $N$.leafNbr $>nbr$}
    \STATE $sid$ = $promoteiSAX(iSAX(N),SAX(q),csl(N))$
    \STATE $N$ = $N.routingtable[sid]$
\ENDWHILE
\STATE sort $N$'s siblings according to lower bound distance into list $l$
\WHILE{number of searched nodes $<nbr$}
    \STATE $N_c$ = pop the head node of $l$
    \STATE fetch all nodes rooted at $N_c$ and search the series inside
\ENDWHILE
\RETURN $k$NN among the visited series
\end{algorithmic}
} 
\end{algorithm}

\subsection{Updates}
\label{sec:update}
As a fully functional index, Dumpy also supports updates (insertion and deletion) besides bulk loading.
For insertion, we first insert the series to the target leaf node.
If the leaf overflows, we read all the SAX words inside and follow the same workflow of index building.
When the query series falls into a full pack, we extract the target leaf node in the pack and redo the node packing for the siblings after a large number of such extractions.
These operations are very fast since these small nodes usually cover a small number of series.

The deletion is almost the same as the iSAX-index family~\cite{isax,ads}.
In particular, we mark the deleted data series in the corresponding leaf via a bit-vector and further insertions can exploit the space occupied by the deleted series while queries ignore these entries.
When a node is empty, we free the occupied space.
The only difference for Dumpy is to update the routing table.

\subsection{Complexity Analysis}
\label{sec:theory}
In this section, we first analyze the time complexity of index building and querying, and then analyze the space complexity.

\noindent{\textbf{[Time complexity]}}
As a disk-based index, the time cost of Dumpy depends on both in-core complexity and disk accesses.
In the following, we first discuss the theoretical time cost in index building and then querying.

The complexity of Dumpy index building could be summed over sub-modules. 
In the adaptive split algorithm, let node $N$ is to be split, and $|csl(N)|$ is between $\lambda_{min}$ and $\lambda_{max}$ according to Equation~\ref{equ:fill}. 
Then the cost of computing the variance of each segment, the base distribution and the routing target is $O(wc_N)$. 
The calling number of the $calcDist$ function is $\binom{w}{\lambda_{max}}*2^{w} + \sum_{i=\lambda_{min}}^{\lambda_{max}-1}\binom{w}{i}2^{\lambda_{i+1}}$=$O(2^w)$, where the first term corresponds to evaluating all possible plans of the max fanout, i.e., using
$\lambda_{max}$ segments, and the second term to evaluating all possible plans of smaller fanouts.
In the leaf node packing algorithm, given that the final pack number is $np$, the time complexity of node packing is $O(2^{\lambda_N}*np)$.
In summary, the total in-core complexity is $O(\sum_N(wc_N + 2^w + 2^{\lambda_N}*np))$.

Random disk writes can have a significant cost when building Dumpy 
(cf. Figure~\ref{fig:pipeline}, Stage 5).
Assume the number of data series in the database is $|db|$, the number of leaf nodes is $n_l$ and the memory buffer can contain $B$ series.
In each batch, Dumpy generates $n_l$ random writes at most, and in total, $O(\frac{|db|}{B}*n_l)$ random writes for the whole index building.

For querying, the approximate search goes down a single path from the root node to a target leaf node. 
Let the length of this path be $|p|$; then the cost is $O(|p|w)$.
The I/O cost is a single disk read of size $O(th)$.
Compared to the iSAX indexes (with binary fanout), the length $|p|$ of the Dumpy path is $2/\overline{\lambda}$x smaller, where $\overline{\lambda}$ denotes Dumpy's average fanout.
In addition to the target leaf node search cost, the complexity of the exact search comprises of $O((1-pr)*n_l)$ random disk reads of size $O(th)$, where $pr$ is the pruning ratio, and $O(w*n_{total}\log n_{total})$ in-core calculations, where $n_{total}$ is the total number of nodes.

In practice, Dumpy is a more compact index (smaller $n_l$ and $n_{total}$ values) than other SAX-based indexes (cf. Section~\ref{sec:expr-build}), and therefore, faster in both building and querying times.

\noindent{\textbf{[Space complexity]}}
The space occupied by Dumpy (in addition to the raw data size) is as follows. 
The SAX words are persisted on disk, 
occupying $\lceil wb*|db| / 8 \rceil$ bytes.
The internal nodes of the index store the routing table, the iSAX word, and the list of segments used in the split (i.e., the chosen segments), for a total of $\sum_{N}(8*2^{\lambda_{N}} + wb/8 + \lambda_{N})$ bytes. 
The leaf nodes store a single iSAX word summary, for a total of $n_l*(wb/8)$
bytes.
Since the number of nodes is small, Dumpy introduces very little additional storage in practice.

\begin{figure*}[tb]
  \subfigure[DumpyOS: Index building workflow]{
\label{fig:DumpyOS-workflow} 
\includegraphics[width=0.56\linewidth]{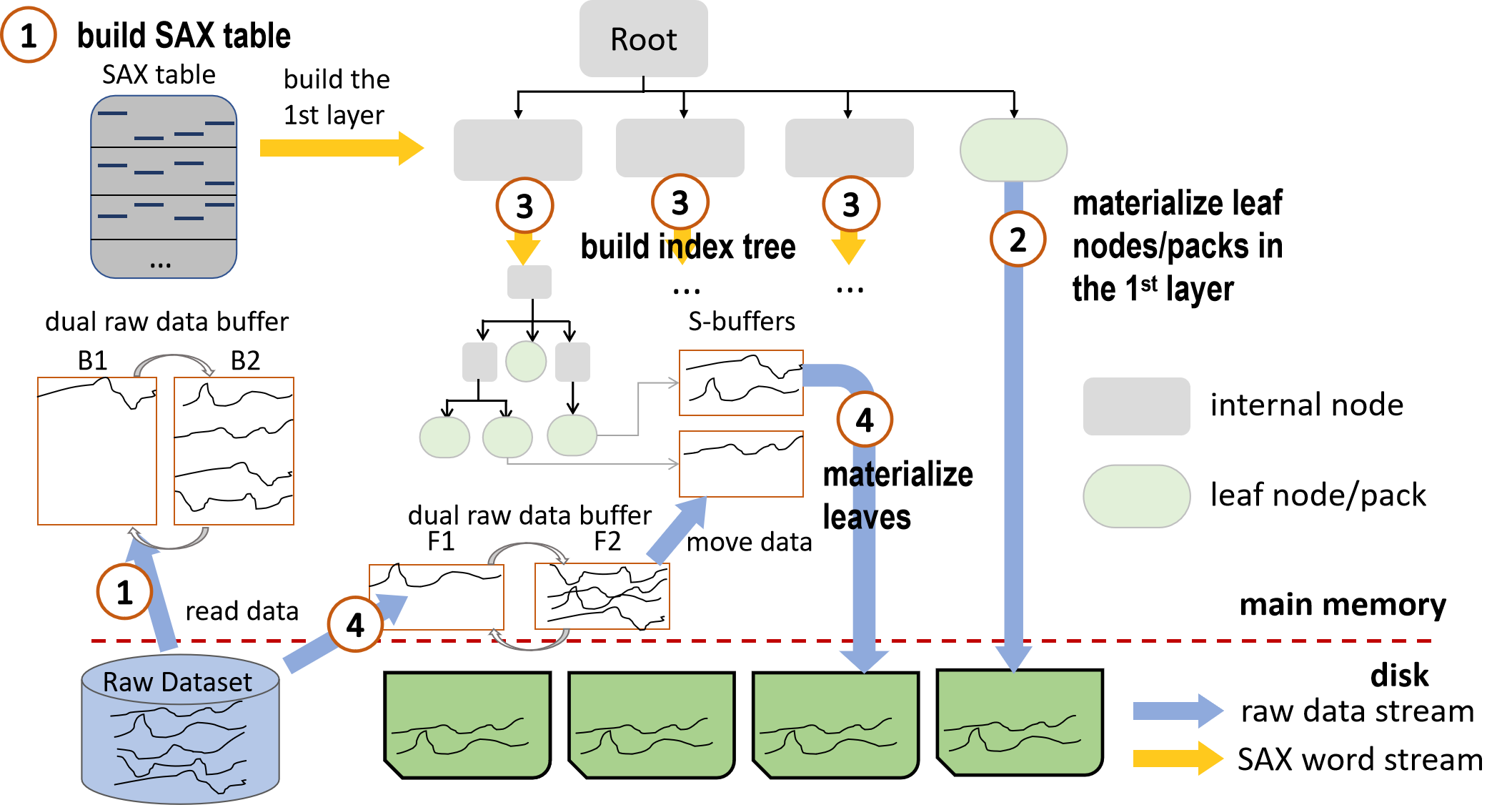}
}
  \subfigure[DumpyOS: Index building timeline of  (time scale does not reflect reality)]{
\label{fig:DumpyOS-timeline} 
\includegraphics[width=0.42\linewidth]{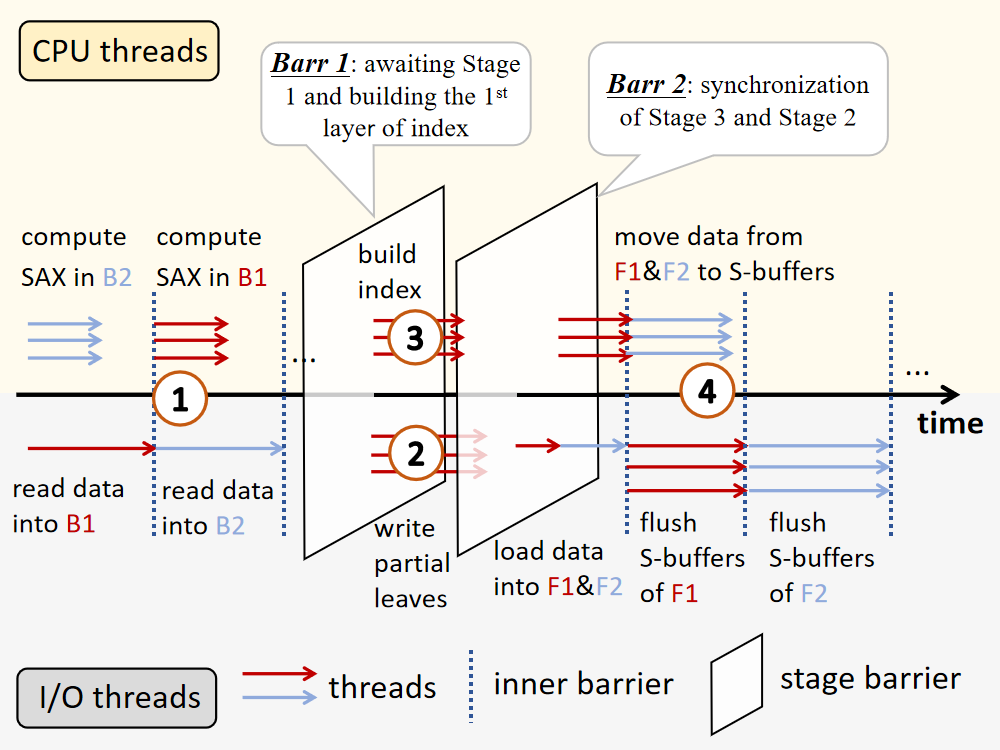}
}
  \caption{{Illustration of building DumpyOS. The inner barriers are inside a loop within a stage and we only show one iteration. The stage barrier is used for synchronization between stages.}}
  \label{fig:DumpyOS-build-illustrate}
\end{figure*}

\section{Dumpy-Fuzzy}
\label{sec:Dumpy-f}
As partition-based indexes, data series indexes also suffer from the so-called \emph{boundary issue} in approximate search~\cite{nsg,spann}.
That is, the data series located near the boundary of a query's resident node 
are also good candidates, but cannot be considered in approximate search since they may be located in different subtrees.
To overcome this problem, we propose a variant of Dumpy, named Dumpy-Fuzzy, that views the static partition boundary (i.e., the SAX breakpoints) as a range (fuzzy boundary) and places the series lying on this range into the nodes of both sides.
Dumpy-Fuzzy further improves the approximate search accuracy compared with Dumpy at the expense of a small overhead on index building and disk space. 

Specifically, Dumpy-Fuzzy adds a duplication procedure after splitting.
For each newly-generated internal node $N$, it checks the series 
lying on $N$'s neighboring nodes (i.e., the nodes whose $sid$ is $1$-bit different from $N$) and duplicates the series near the boundary into itself.
We introduce a hyper-parameter $f \in (0,1)$, the fuzzy boundary ranges regarding the original node ranges, to control which series is qualified to be duplicated.
In addition, duplication also applies after leaf node packing.
The series near the boundaries of a leaf pack can also be placed redundantly into the pack 
in the same way as above.
we ensure that no additional split will be introduced in this procedure (i.e., the leaf pack will not overflow).

Note that Dumpy-Fuzzy does not damage Dumpy's pruning power for exact search.
Duplicated series do not change the iSAX words of nodes or packs. 
Hence, the lower bound calculations are kept the same.
Therefore, without violating the pruning-based exact search, Dumpy-Fuzzy improves the approximate search accuracy
by examining more promising candidates.


\section{DumpyOS}
\label{sec:DumpyOS}
In this section, we introduce DumpyOS, which extends Dumpy in a parallel way to maximize its performance under modern multi-core architectures and NVMe devices.
We first introduce some preliminary materials about the characteristics of NVMe SSD devices, and present an overview of our methods in Section~\ref{sec:preliminary}.
Then, we describe the concurrent index building workflow in Section~\ref{sec:DumpyOS-build}, and the novel parallel pruning-based query answering techniques in Section~\ref{sec:DumpyOS-query}.

\subsection{Overview}
\label{sec:preliminary}
In the following, we briefly describe the characteristics of NVMe SSDs, and summarize key rules to better exploit them.
Then we introduce the rationale of DumpyOS based on these key rules.

\noindent\textbf{Rule 1: Issue large batches of parallel I/Os to ensure the internal parallelization is fully exploited~\cite{ssd-vldbj,ssd-tos}}.
The most prominent feature of SSD is the \emph{internal parallelism}.
The storage unit of SSD (i.e., the flash memory) is organized in a highly-hierarchically manner to maximize I/O concurrency~\cite{ssd-sim}.
The storage units are organized into different levels 
(e.g., page, block, die, and channel from bottom to top).
The upper levels can operate independently to serve various requests simultaneously, while the lower levels can only work in parallel if they are executing identical commands.

\noindent\textbf{Rule 2: Separate read and write operations.}
Although parallel read and write operations are beneficial to the exploitation of SSDs, the interference between reads and writes may cause access conflicts and hence hurt the parallelism and I/O performance~\cite{ssd-confict}.

\noindent\textbf{Rule 3: Buffer small random writes into large sequential writes.}
Another peculiarity of flash memory is garbage collection.
Given that flash memory must be erased before it can be rewritten, with erasures at the block granularity and writes at the page granularity, the original data need to be copied to another block, known as the write amplification on SSD~\cite{ssd-wikipedia,ssd-understand}.
Writes smaller than a single page actually perform a read-modify-write operation on SSD, and result in one invalid page.
This leads to not only more unnecessary reads and writes, but also to more frequent garbage collection~\cite{ssd-smallwrite,ssd-smallwrite2}.

Based on these rules, our proposed method, DumpyOS, divides the operations in index building and querying into two separate parts, i.e., I/O and CPU operations, and then adopts different parallel techniques to accelerate each one of these two parts.
Furthermore, we design a dedicated workflow to overlap CPU calculation time with I/O time by interleaved threads scheduling.

\subsection{Concurrent Index Building}
\label{sec:DumpyOS-build}
To better exploit the parallelism of modern hardware, we design a new concurrent workflow for DumpyOS, shown in Figure~\ref{fig:DumpyOS-build-illustrate}. 

The DumpyOS workflow consists of four stages.
In Stage~1, we build the SAX table with a dual raw data buffer (one reads data, while the other computes the SAX words~\cite{paris+}), and then we build the first layer of the index tree~\footnote{Recall that the root node always chooses all $w$ segments without costly adaptive splitting.}.
After synchronization of these threads (Barr~1), Stages~2 and~3 are executed in parallel.
Stage~2 reads raw data in batches, routes them into leaf nodes, or packs in the first layer, and materializes these leaves to SSD.
At the same time, in Stage~3, we employ a group of threads as index-building workers to grow the index tree from internal nodes of the first layer (one worker for each subtree).
Since there is a significant size difference between different internal nodes (cf. Figure~\ref{fig:dist-toplayersplits}), we schedule the threads in a dynamic manner.
Each worker fetches an internal node, conducts adaptive splits and node packing, and then repeats this process until all the subtrees are built. 
Once all threads from the two stages return (Barr~2), Stage~4 starts.
Similar to Stage 2, Stage~4 reads raw data and materializes other leaves (below the first layer). 

Note that Stage~2 is additionally incurred to materialize the leaves in the first layer.
This stage involves an I/O-bound process that utilizes the available I/O time during index structure construction (Stage~3).
Although one extra pass of sequential read is introduced, it is very fast compared to leaf materialization (random writes).
This stage has the additional advantage that it reduces the burden of Stage~4.


\noindent{\textbf{[Speedup materialization with parallel I/Os]}} 
Since data movements on secondary storage are usually the bottleneck in index construction~\cite{paris+,dumpy}, improving the I/O throughput with SSDs becomes a promising direction to accelerate the leaf materialization process.
Recall that the materialization process consists of three steps: (i) read a series from the raw dataset, (ii) find the leaf to which the series belongs, and (iii) append the data series to the leaf file once the memory is full.

To reduce the number of I/Os, a buffer pool is usually prepared to store the data read from the dataset.
However, such a serial design does not fully use the SSD's internal parallelism.
Following \textbf{Rule~1}, a naive I/O-parallel method is to divide the buffer pool into several pieces and assign each of them to a single thread, which repeats the three steps.
Although this method achieves I/O parallelization, it mixes read and write requests issued from different threads, and thus, generates severe I/O conflicts among threads, leading to a degradation of the SSD performance (\textbf{Rule~2}).



Consequently, we can set barriers between reads and writes in order to isolate them.
That is, we parallelize each one of the three steps with barriers between them.
Specifically, we (sequentially) read a batch of data into a unified buffer pool (rather than small buffer pieces), find the target leaf nodes of the series, and finally write these leaves one by one on the SSD.
For the second step, we can partition the buffer in equally-sized partitions, and assign these partitions to a group of threads (one for each partition).
The threads find the target leaf for each series, and a concurrent hash map (where the key is a leaf and the value is a set of pointers to the series in the buffer pool) can be used to collect the results.
As for the third step, the prepared leaves can be materialized in parallel by threads without any conflicts\footnote{There are many ways to implement parallel I/O requests (e.g., asynchronous I/O with io\_uring~\cite{iouring} or libaio~\cite{libaio}). For simplicity and clarity, we use multi threads to issue I/O requests in this paper.}.

The writing process, which is still a bottleneck, can be further optimized.
In the algorithm above, we issue one write request for a single series ($\approx$1KB for 256-length), leading to very small writes and performance degradation (\textbf{Rule~3}).
Therefore, we optimize it by buffering small writes.
The second step is modified to move the full data series (rather than pointers) to the buffers of target leaves, named S-buffers.
Finally, the third step can directly flush S-buffers to the SSD (i.e., one I/O request for each buffer).
In this way, the internal parallelism of modern SSDs is fully leveraged and the throughput is significantly improved.
However, when we move data to the S-buffers of leaves (the second step described above), an additional pass of data-moving operations is introduced in memory, which incurs a substantial overhead.
In the following, we explain how we address this by interleaved thread scheduling.

\noindent{\textbf{[Mask CPU time with interleaved threads]}} 
As shown in Figure~\ref{fig:DumpyOS-build-illustrate} (Stage~4), we 
divide the unified memory buffer into two parts (F1 and F2).
In each cycle, we first read data to fill F1.
Next, a group of CPU-bound threads is initiated as routing workers to move data from F1 to the corresponding S-buffers.
In the meanwhile, another I/O-bound thread reads data into F2 (interleaving read and computing).
We set an inner barrier here to await the completion of these threads.
Then, the routing workers start to move data from F2 to the S-buffers, while we start a group of I/O threads to flush the prepared S-buffers originating from F1 (interleaving writes and computing).
Another inner barrier is set here to synchronize these threads.
After that, the threads flush the S-buffers from F2. 
This process is repeated until a full pass of the raw dataset is completed.
In this way, we mask the extra CPU time and always keep the SSD busy with maximal throughput.

\begin{figure}[tb]
  \includegraphics[width=1\linewidth]{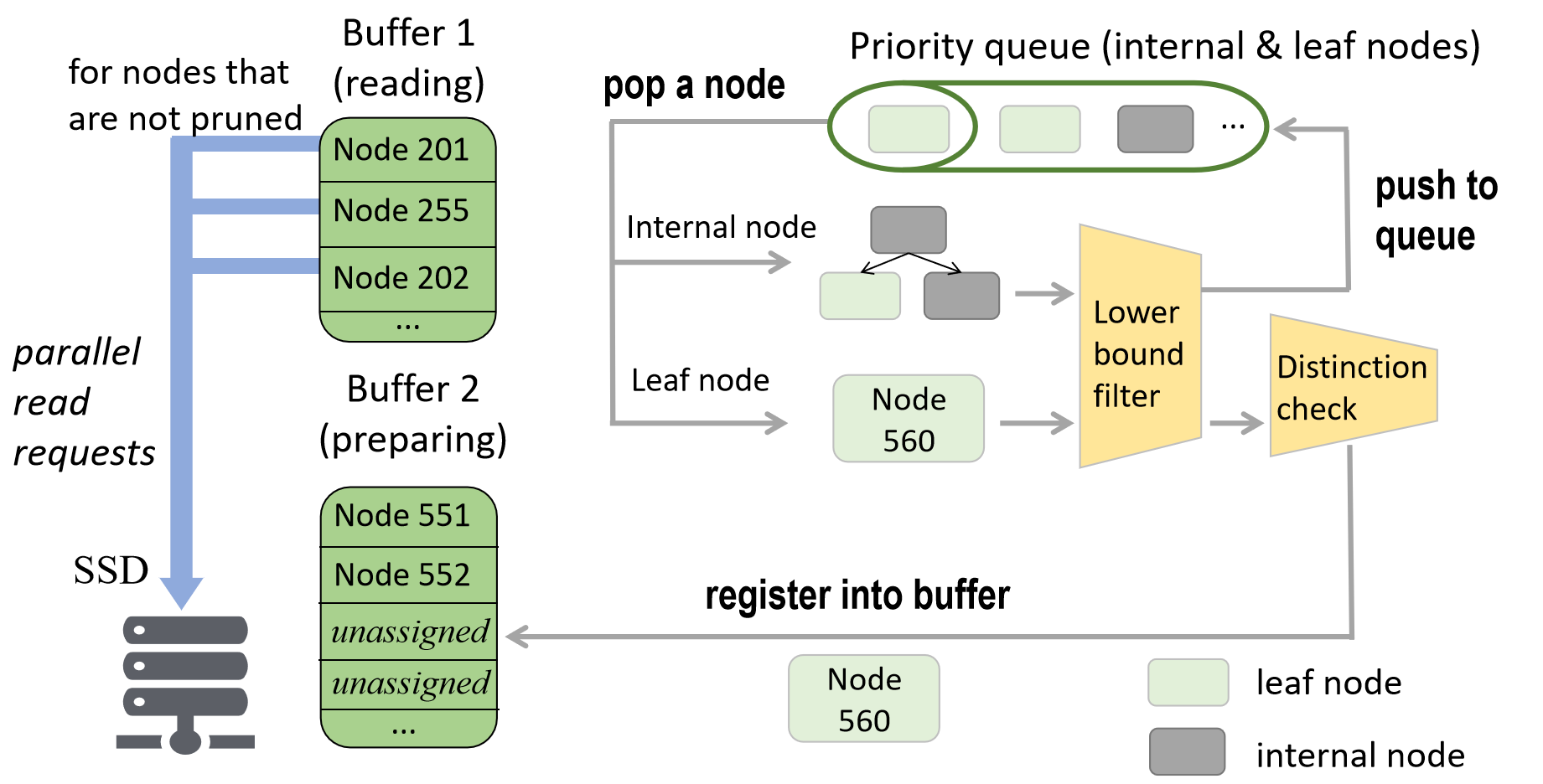}
  \caption{An illustration of parallel pruning-based query algorithm of DumpyOS. It can also work on Dumpy on HDD to mask the CPU time, but the I/O time is still the bottleneck.}
  \label{fig:p-query}
\end{figure}
\subsection{Parallel Pruning-based Query Answering}
\label{sec:DumpyOS-query}

In this subsection, we focus on accelerating the pruning-based query algorithm based on modern parallel hardware.
SOTA parallel query algorithms like PARIS+~\cite{paris+} and Coconut~\cite{coconut} achieve speedup by serially scanning the dataset.
However, this search style has three limitations: 
(1) it cannot serve approximate queries like $ng$- and $\delta\epsilon$-search~\cite{hydra2};
(2) it is impractical to always maintain a sequential data layout and aligned in-memory SAX table for a dynamic workload in a real situation; and 
(3) it requires additional space in the memory and secondary storage to store the SAX table and dataset, respectively, which limits scalability.

In our case, we decided to optimize the classical pruning-based query algorithm to achieve higher performance, while avoiding these limitations.
Our core idea is shown in Figure~\ref{fig:p-query}.
We divide the threads into two groups, loading and computing workers respectively.
Each group gets allocated a memory buffer at a time that contains $\eta$ slots to accommodate leaf nodes or packs ($\eta$ is a user-defined parameter depending on the parallelism of the SSD).
First, we initialize a priority queue with the root node (ordered by the lower bound distance in ascent).
Then one computing worker takes a node from the priority queue and if it is an internal node, then we perform a qualification check on its children by comparing their lower bound distances to the query.
Qualified children nodes are pushed into the queue.
If the node is a leaf that has not been visited before, it will be registered into a slot of the current buffer (i.e., Preparing Buffer), along with the file position of this leaf.
Once the buffer is filled with $\eta$ leaves, the computing workers will wait for loading workers, who are meanwhile reading raw data from data files into the other buffer (i.e., Reading Buffer), which has been filled with candidate leaves beforehand.
In this way, we can parallelize the read requests in querying to increase the I/O performance (\textbf{Rule~1}).
Once loading workers finish, we exchange the two buffers, and immediately let loading workers read data for the new buffer.
Consequently, we wake the computing workers to compute the distances of the data-ready buffer in parallel with SIMD techniques~\cite{messi}, and update current BSF answers.
Finally, the buffer is cleared and the above process is repeated until the queue head violates the condition of the lower bound filter. 
When this happens, it is impossible to produce better answers from the nodes of the queue; for the nodes already in the buffer, we will still fetch the data from SSDs and compute distances for them before terminating the algorithm. 
Overall, in our algorithm, both CPU computations and I/O requests are fully parallelized, enjoying the performance improvements stemming from modern hardware.
\begin{figure*}[tb]
\begin{minipage}{.34\linewidth}
    \includegraphics[width=\linewidth]{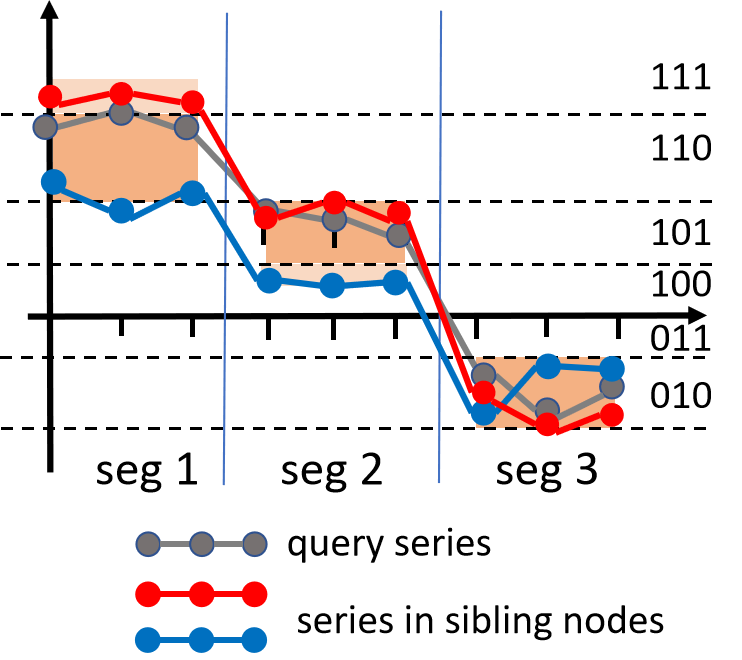}
    \caption{Illustration of the downside of Dumpy-Fuzzy. The red series is closer to the query series than the blue one.}
    \label{fig:dumpyf-problems}
\end{minipage}%
\hspace{3mm}
\begin{minipage}{.66\linewidth}
    \includegraphics[width=\linewidth]{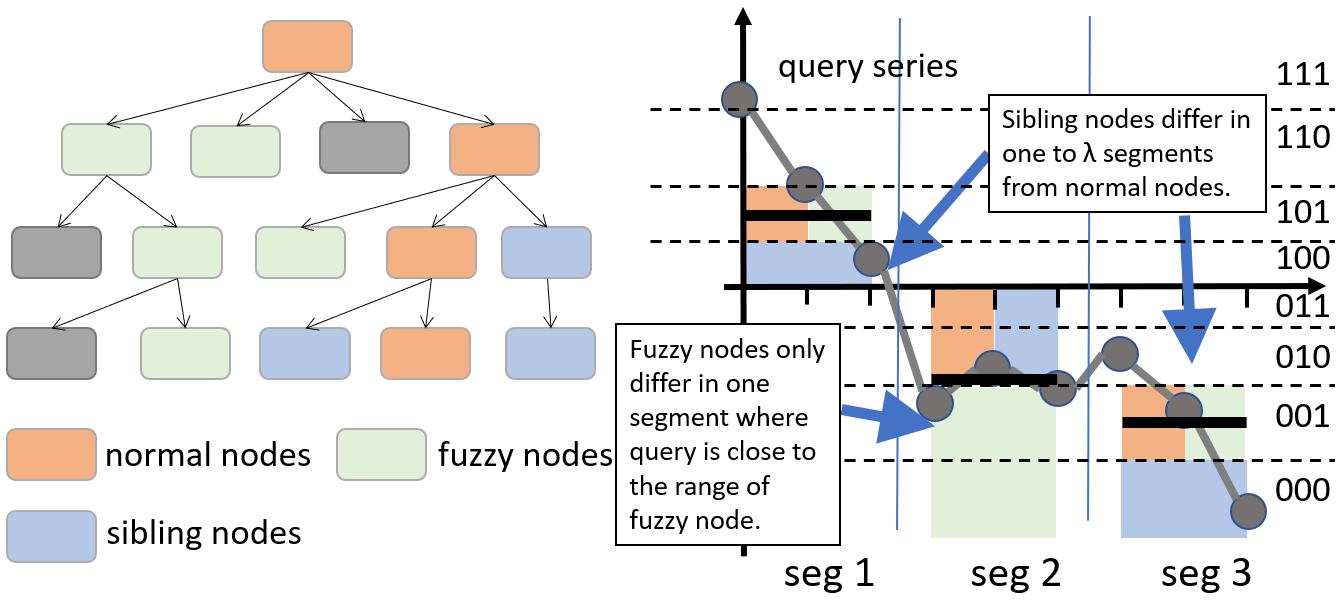}
    \caption{Extended approximate search with DumpyOS-F. 
 (left)Three kinds of nodes visited during searching. (right) Illustration of fuzzy and sibling nodes; cells marked in two colors indicate that they are occupied by two nodes.}
    \label{fig:dumpy-vf-illustate}
\end{minipage}%
\vspace{-3mm}
\end{figure*}


\section{DumpyOS-F}
\label{sec:dumpy-vf}
In this section, we introduce an enhanced algorithm for extended approximate search, DumpyOS-F.
Compared to Dumpy-Fuzzy, the most prominent characteristic of DumpyOS-F is that it does not need any modification of Dumpy's structure, and thus no additional space cost.
At the same time, it provides more accurate results.

While Dumpy-Fuzzy mitigates the boundary issue by physically duplicating series into neighboring subtrees or nodes, it still has three limitations that hinder the improvement of search accuracy: 

\noindent(1)
The number of duplicated series is limited to prevent excessive dilation of the original index structure.
The duplication in a leaf node or pack is restricted to avoid overflow by abandoning checking other neighboring series despite a higher similarity.
The selection of duplicated series is random since it is not possible to know which series is closer to the query during indexing.
For example, in Figure~\ref{fig:dumpyf-problems}, we cannot decide whether it would be better to duplicate the red, or the blue series into the node, without knowing the query series. 
This leads to possible loss of promising candidate series.

\noindent(2)
The similarity between duplicated series and the query is limited.
As shown in Figure~\ref{fig:dumpyf-problems}, consider that the capacity of the node is sufficient, and thus both the red and blue series are contained in this node.
Given that the blue series is not as close to the grey query series as the red one, duplicating the blue series does not provide any benefit for this query.
Nevertheless, Dumpy-Fuzzy will duplicate the blue series, because of its proximity to the middle segment (seg 2) of the node.

\noindent(3)
The same series may be checked multiple times.
Since the duplicated series come from the sibling nodes, given our extended search Algorithm~\ref{alg:search}, it is possible to obtain the same series during searching (e.g., the red series exists in two nodes and can be visited twice), making the duplication meaningless in this case.

In summary, since Dumpy-Fuzzy statically selects similar series for a node to duplicate during indexing, it cannot accurately identify the series that are similar to a given query series, yet lying in different nodes.


In DumpyOS-F, we determine the candidate series during querying rather than indexing.
DumpyOS-F estimates the distance between the query series and sibling nodes on-the-fly, and further prioritizes these nodes.
That is, we only check the series in the fuzzy boundary that are close to the query itself rather than the node where the query lies.
As a result, DumpyOS-F is more accurate and efficient in approximate search, and most importantly, it works directly on Dumpy's index structure. 
Therefore, users do not need to build a specialized index for approximate search like Dumpy-Fuzzy.

As shown in Figure~\ref{fig:dumpy-vf-illustate} (left), DumpyOS-F accesses three kinds of nodes in search: normal, fuzzy, and sibling nodes.
Normal nodes are the nodes we visit when routing a query series from the root to a leaf node.
That is, the query series is located in the range of normal nodes.
At each layer, we mark the nodes that are close to the query series as fuzzy nodes.
Specifically, we compute the distance between the PAA value of the query series and the corresponding breakpoint at each segment, and check whether the query series is located in the fuzzy boundary of a sibling node (according to parameter $f$).
If so, the sibling node is marked as a fuzzy node.
For example, in Figure~\ref{fig:dumpy-vf-illustate} (right), the second segment of the query series PAA is very close to the breakpoint. 
Thus, among all sibling nodes, we select the one that only differs from the normal node (marked in orange) on the \emph{second} segment: this selected node becomes the fuzzy node (marked in green).
Finally, some other siblings of the normal nodes are marked as sibling nodes that will also be visited (as described in Algorithm~\ref{alg:search}), which are shown in blue in Figure~\ref{fig:dumpy-vf-illustate}.

The pseudocodes of DumpyOS-F are shown in Algorithm~\ref{alg:dumpy-vf}.
In the first step (lines 1-16), we route the query series to the target leaf, while collecting the fuzzy nodes at each layer into a priority queue according to their estimated distances in ascent (lines 6-12).
Since a pack containing multiple small leaves may be put into the queue many times, among these entries, we only keep the one with the smallest estimated distance (line 15).
In the second step (lines 14-21), we search the target leaf and then pop the fuzzy nodes from the priority queue (line 17).
If the popped fuzzy node is an internal node, we called the adapted routing algorithm (Algorithm~\ref{alg:adapted-routing}) to obtain the nearest leaf to the query series and search it (line 19).
If the popped node is a leaf, we will directly search the series inside (line 20).
Finally, if the number of accessed nodes is still within budget, we search sibling nodes in a bottom-up fashion by calling Algorithm~\ref{alg:search}.

The adapted routing algorithm is shown in Algorithm~\ref{alg:adapted-routing}.
In line 1, we first identify the special segment, where the fuzzy node's iSAX label is not a prefix of the query's corresponding SAX label, and use bit $b$=$1$ (line 2) to record the (series inside the) node is smaller than the query in this segment, $b$=$0$ otherwise.
For example, consider the green fuzzy node in Figure~\ref{fig:dumpy-vf-illustate} (right) as $N$.
Then the second segment will be identified since the node's second iSAX symbol $00$ is not a prefix of the query's $010$.
$b$ is set to $1$ as $N$ is below the query series in this segment.
Then, if $N$ is split on the second segment, we always route the query series to the upper child (i.e. whose second iSAX symbol is 00\textbf{1}), because it is closer to the query than the lower child.
To achieve this, we slightly modify the iSAX promotion algorithm as shown in Function PromoteiSAXF, where we always set the bit of the special segment to the value $b$.
Since at each layer we have at most $\lambda$ fuzzy nodes and in the worst case they are all internal nodes, then the extra time cost is $O(|p|*\overline{\lambda})$, which is negligible compared to node loading and distance calculation.

\begin{algorithm}[tb]
\caption{DumpyOS-F} 
\label{alg:dumpy-vf}
{
\footnotesize
\begin{algorithmic}[1]
\REQUIRE root node $N_r$, node number $nbr$, query series $q$, fuzzy boundary parameter $f$
\STATE node $N$ = $N_r$
\STATE initialize an empty priority queue $pq$
\WHILE{$N$ is an internal node}
    \STATE $sid$ = $promoteiSAX(iSAX(N), SAX(q), csl(N))$
    \STATE $N_c$ = $N.routingtable[sid]$
    \FOR{each segment $seg$ in $csl(N)$}
        \STATE $bp$ = next breakpoint in $seg$
        \STATE $r$ = value range of $iSAX(N_c)[seg]$
        \STATE $es\_dis$ = $|PAA(q)[seg] - bp|$
        \IF{$es\_dis < f*r$}
            \STATE $sib\_id$ = flip the bit at $seg$ of $sid$
            \STATE push $<N.routingtable[sib\_id], es\_dis>$ into $pq$
        \ENDIF
    \ENDFOR
    \STATE $N$ = $N_c$
\ENDWHILE
\STATE search node $N$
\STATE remove the entries in $pq$ that point to the same node but have larger distances
\WHILE{$pq$ is not empty and $nbr > 1$}
    \STATE $<N_c, \_>$ = pop the first element from $pq$
    \IF{$N_c$ is an internal node}
        \STATE $N_c$ = $AdaptedRouting(N_c, q)$
    \ENDIF
    \STATE search node $N_c$
    \STATE $nbr -=1$
\ENDWHILE
\STATE $ExtendedApproximateSearch(N_r, nbr, q)$ while skipping the visited nodes
\RETURN $k$NN among the visited series
\end{algorithmic}
} 
\end{algorithm}

\begin{algorithm}[tb]
\caption{Adapted Routing} 
\footnotesize
\label{alg:adapted-routing}
{
\begin{algorithmic}[1]
\REQUIRE fuzzy node $N$, query series $q$
\STATE $seg$ = the segment where $iSAX(N)[seg]$ is not a prefix of $SAX(q)$
\STATE $b$ = the inversion of the last bit at $iSAX(N)[seg]$
\WHILE{$N$ is an internal node}
    \STATE $sid$=$PromoteiSAXF$($iSAX(N)$,$SAX(q)$,$csl(N)$,$seg$,$b$)
    \STATE $N$ = $N.routingtable[sid]$
\ENDWHILE
\RETURN $N$

\hrulefill
\renewcommand{\algorithmicrequire}{\textbf{Function}}
\REQUIRE PromoteiSAXF(iSAX word $isax$, SAX word $sax$, chosen segments list $csl$, segment $s$, fixed symbol $b$)
\STATE $sid$ = 0
\FOR{each segment $seg$ in $csl$}
    \STATE $nb$ = $len(isax[seg])$
    \IF{$seg = s$}
        \STATE $sid$ = ($sid << 1$) + $b$
    \ELSE
        \STATE $sid$ = ($sid << 1$) + the $(nb+1)$-th bit of $sax[seg]$

    \ENDIF
\ENDFOR
\RETURN $sid$
\end{algorithmic}
} 
\end{algorithm}

\section{Experiments}
\label{sec:expr}
\noindent\textbf{[Environment]} Experiments were conducted on an Intel Core(R) i9-10900K 2.80GHz 10-core CPU with 4*32GB 2400MHz main memory, running Windows Subsystem of Linux (Ubuntu Linux 20.04 LTS). The machine has a Samsung PCIe 2TB SSD (default), and a Seagate SATAIII 7200RPM 2TB HDD. 
Our codes are available at https://github.com/DSM-fudan/DumpyOS.

\noindent\textbf{[Datasets]} 
We use one synthetic and three real datasets. All series are z-normalized before indexing and querying. 
In each dataset, we prepare 200 queries that are not part of the dataset of varying hardness~\cite{hardness}, and obtain the ground truth kNN results using brute-force search.
\textbf{Rand} is a synthetic dataset, generated as cumulative sums of random walk steps following $N(0,1)$. 
It has been extensively used in the existing works~\cite{dft,DBLP:journals/vldb/ZoumpatianosLIP18,hydra1,hydra2}. 
We generate 50-800 million Rand series of different lengths (50GB-800GB). 
\textbf{DNA}~\cite{dna} is a real dataset collected from DNA sequences of two plants, Allium sativum and Taxus wallichiana.
It comprises 26 million data series of length 1024 ($\sim$113GB).
The second real dataset, \textbf{ECG} (Electrocardiography), is extracted from the MIMIC-III Waveform Database~\cite{ecg}.
It contains over 97 million series of length 320 ($\sim$117GB), sampled at 125Hz from 6146 ICU patients.
The last real dataset, \textbf{Deep}~\cite{deep}, comprises 1 billion vectors of size 96, extracted from the last CNN (convolutional neural network) layers of images.

\noindent\textbf{[Algorithms]} 
In iSAX-index family, we take \textbf{iSAX2+} as the SOTA binary structure~\cite{hydra2}.
We also implement a stand-alone version of \textbf{TARDIS} as the SOTA full-ary structure and use 100\% sampling percent.
\textbf{DSTree}~\cite{ds-tree} 
is also included as one of the SOTA data series indexes~\cite{hydra2}.
For simplicity, Dumpy-Fuzzy with parameter $f$ is abbreviated as \textbf{Dumpy-$f$}. 
To evaluate the quality of these indexes, we implement extended approximate search, as well as pruning-based search on them.
We also include PARIS+~\cite{paris+} as the SOTA parallel (serial-scan-based) disk index to be compared to DumpyOS.


All the codes are open-source, implemented in C and C++, compiled by g++ 9.4.0 with -O3 optimization. 

\noindent\textbf{[Parameters]} We set the number of segments $w$=16, SAX cardinality $c$=64 (i.e., $b$=8), $\alpha$=0.2, and the leaf size threshold $th=10000$. 
The replication factor of each series in Dumpy-$f$ is set to at most 3, and $f$ is set to 10 for ECG and Deep and 30 for Rand and DNA.
The discussion about the influence and the selection of parameters can be found in \cite{dumpy}.
The memory buffer size for index building is set to 4GB unless specified.
The default number of threads for DumpyOS is set to 5 for index building and $8$ for querying.
The queue size $\eta$ is set to 24 for our SSD.

\noindent\textbf{[Measures]}
Similar with other works~\cite{hydra2,hd-index,new-pq}, we use 
Mean Average Precision (MAP)~\cite{map} as the accuracy measure, which is defined as the mean value of AP on a group of queries. For query $s_q$, AP equals to
$\frac{1}{k}\sum_{i=1}^{k}P(s_q\\,i)*rel(i)$, where $P(s_q,i)$ is the ratio of true neighbors among the top-$i$ nearest  results and $rel(i)$ is 1 if the $i$-th nearest result is the true exact $k$NN result and 0 otherwise. It can be proved that MAP is equivalent to the average recall rate when the returned results are sorted by the actual distances.
Another similarity measure we use is the average error ratio which measures the difference between approximate and exact results, commonly used in approximate search ~\cite{hd-index,tardis}, and defined as $\frac{1}{k}\sum_{i=1}^{k}\frac{dist(a_i,s_q)}{dist(r_i,s_q)}$.
We measure both ED and DTW, where the DTW warping window size is set to $10\%$ of the series length as a common setting~\cite{messi,dtw}.

\begin{figure}[tb]
\includegraphics[width=\linewidth]{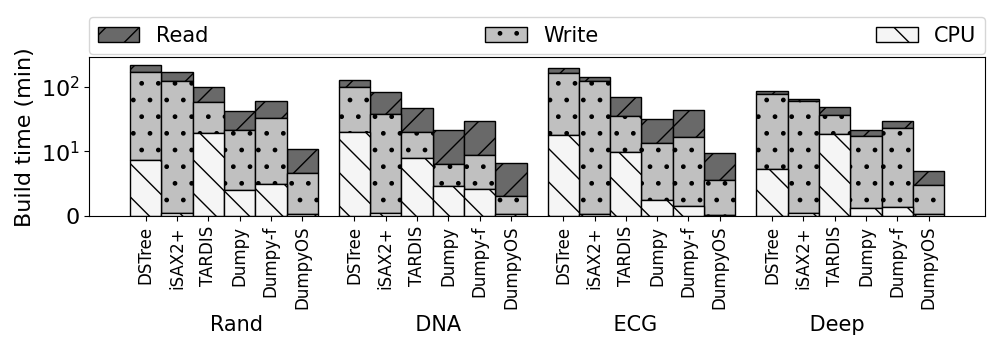}
\caption{Index building time across four datasets on SSD.}
\label{fig:cons} 
\vspace{-2mm}
\end{figure}

\vspace{-5mm}
\begin{table}[tb]
\label{tab1}
\centering
\scriptsize
	\begin{threeparttable}
	\caption{Index structure statistics}
	\begin{tabular}{c|c|c|c|c|c|c}
		\hline
		Data & Method & \#Leaves & \#Nodes  & Ht. &  \thead{Fill\\factor} & \thead{Size\tnote{1}}\\
		\hline\hline
		\multirow{4}{*}{Rand} & iSAX2+ & 73563 & 86945 & 20  &13.59\%  & {16}\\
		\cline{2-7}
		&                 DSTree & 17847 & 35693 & 32  & 56.03\%  & {9}\\
		\cline{2-7}
		&                TARDIS   &8516867  & 8520065 & \textbf{3} & 0.11\% & {732}\\
		\cline{2-7}
		&                \textbf{Dumpy}   & \textbf{14106} &  \textbf{19418} & 7 & \textbf{70.89\%}  & {\textbf{3}}\\
		\hline\hline
		\multirow{4}{*}{DNA}  & iSAX2+ & 42906 & 47885 & 25  & 6.14\%  & {9}\\
		\cline{2-7}
		&         DSTree &  5833& 11665 &  43 & 45.16\%  & {3}\\
		\cline{2-7}
		&         TARDIS   &  1011436 & 1312989  & \textbf{5} & 0.26\% & {278}\\
		\cline{2-7}
		&         \textbf{Dumpy}   & \textbf{4367} &  \textbf{6228} & 9 & \textbf{60.32\%} & {\textbf{1}}\\
		\hline\hline
		\multirow{4}{*}{ECG}  & iSAX2+ & 69786 & 74042 & 9  & 13.98\%  & {14}\\
		\cline{2-7}
		&     DSTree & 20740 & 41479 & 48  & 47.04\% & {13}\\
		\cline{2-7}
		&     TARDIS   & 3178628 &3182368  &\textbf{4}  & 0.33\% & {749}\\
		\cline{2-7}
		&     \textbf{Dumpy}   & \textbf{12112} & \textbf{15050} & 7 & \textbf{80.55\%} & {\textbf{3}} \\
		
		\hline\hline
		\multirow{4}{*}{Deep}  & iSAX2+ & 68096 & 71188 & 8  & 19.08\% & {11}\\
		\cline{2-7}
		&     DSTree & 16324 & 32647 & 33  & 61.26\% & {8}\\
		\cline{2-7}
		&     TARDIS   & 824458  & 827094 & \textbf{3} & 0.27\% & {546}\\
		\cline{2-7}
		&     \textbf{Dumpy}   & \textbf{11590} & \textbf{13664} & 8 & \textbf{86.28\%} & { \textbf{3}} \\
		
		\hline\hline
		
	\end{tabular}
	\begin{tablenotes} 
        \footnotesize   
        {
        \item[1] Size of in-memory index structure only, in MB unit. 
        }
      \end{tablenotes}         
    \end{threeparttable}
     \vspace{-5mm}
\end{table}

\subsection{Index Building}
\label{sec:expr-build}
\subsubsection{Efficiency}
First, we evaluate the index-building efficiency in four datasets on SSD, and the results are shown in Figure~\ref{fig:cons}. 
In all four datasets, Dumpy outperforms the other three methods by a large margin, i.e, $5.3$ times faster than DSTree, $3.8$ times than iSAX2+ and $2.5$ times than TARDIS on average. 
Dumpy-$f$ only incurs small overheads (about \textbf{38\%}) on Dumpy and is considerably faster than DSTree and iSAX2+.
Moreover, DumpyOS is \textbf{3.7} times faster than Dumpy (with 5 threads).
Note that in this figure, the CPU time is only for the part that does not overlap with I/O time.
(A detailed analysis of DumpyOS will be presented in Section~\ref{sec:dumpyos-build}.)


We present the detailed index information in Table 1. Dumpy has the fewest leaf nodes (i.e., the highest fill factor), which verifies the good compactness of Dumpy. The number of leaf nodes of DSTree is slightly larger than Dumpy, and that of iSAX2+ is generally $>$\textbf{3}x more than DSTree.
As for the full-ary structure like TARDIS, it generates millions of leaves and has a low fill factor in the leaves, as analyzed in Section~\ref{sec:trade-off}.
These nodes are further packed into large partitions of capacity 128MB, as the setting of the original paper~\cite{tardis}.
These results verify the space complexity analysis in Section~\ref{sec:workflow}.

\begin{figure}[tb]
\includegraphics[width=0.95\linewidth]{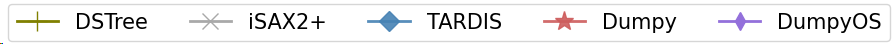}
\subfigure[On data series length]{
  \includegraphics[width=0.47\linewidth]{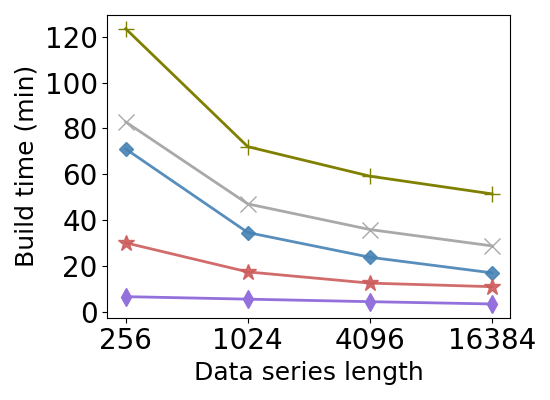}
  \label{fig:cons-length}
}
\subfigure[On dataset size]{
\includegraphics[width=0.47\linewidth]{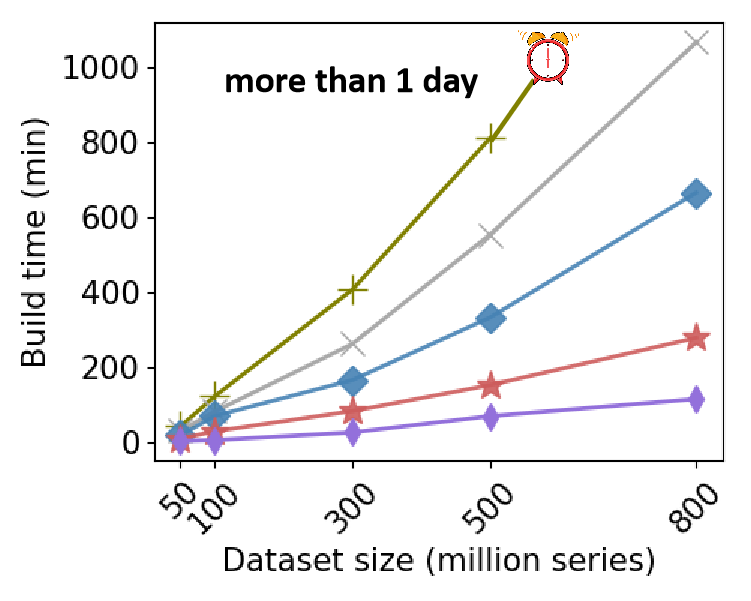}
\label{fig:cons-size}
}
\caption{Index scalability (32GB memory).}
\label{fig:scale} 
\end{figure}

\begin{figure}[tb]
\includegraphics[width=\linewidth]{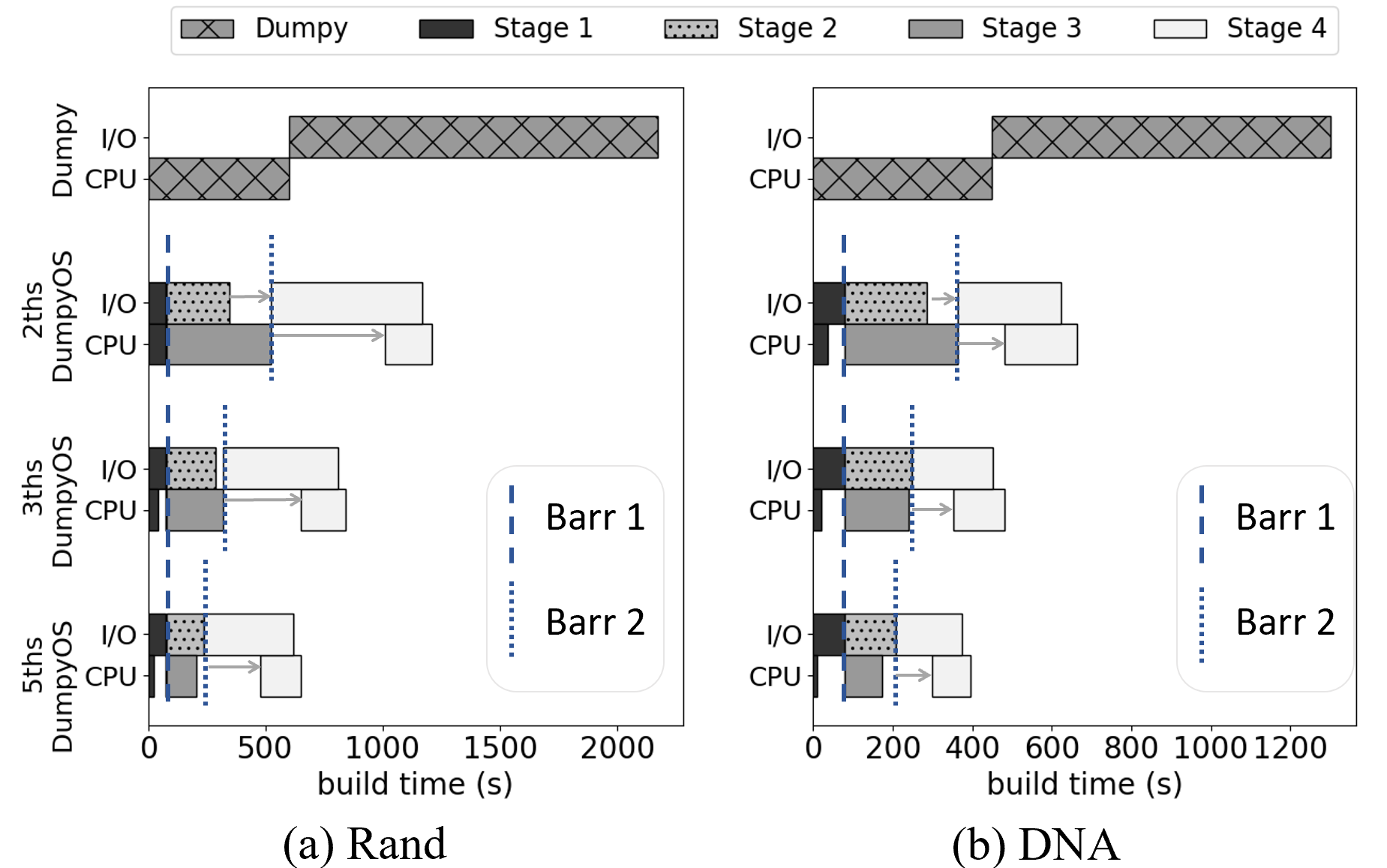}
\caption{Parallel building with DumpyOS with 2, 3, and 5 threads, on two 100GB datasets (Rand and DNA). Dotted lines denote stage barriers. Grey arrows indicate that the corresponding threads are waiting for synchronization (inner barriers), and for clarity, we do not show the detailed timelines between inner barriers. That is, the I/O and the CPU time of each stage are summed as a complete period.}
\label{fig:DumpyOS-build} 
\end{figure}

\subsubsection{Scalability}
Next, we test the scalability in Rand datasets by increasing data size from 50GB to 800GB, and series length from 256 to 16384.
When the dataset size is varied, the series length is kept constant at 256, whereas the dataset size is kept at 100GB when the length is varied, as the same design with the benchmark~\cite{hydra1}.

Figure~\ref{fig:scale} presents the index building time with 32GB memory. 
Dumpy has the best scalability in both cases.
In a linear regression test for the building time and dataset size, Dumpy's coefficient of determination $R^2$ is \textbf{0.99}, verifying its linear growth of the building time.
The reason is that the number of leaves increases linearly as the dataset scales up, indicating a nearly constant average fill factor.
This also supports the complexity analysis.
The performance when varying series length also follows this rule.
Besides that, DumpyOS shows a stable reduction in indexing time as the dataset size increases (about \textbf{3} times faster than Dumpy).

\subsubsection{Parallel Construction with DumpyOS}
\label{sec:dumpyos-build}
In Figure~\ref{fig:DumpyOS-build}, we show the building time breakdown of DumpyOS.
In the first two rows, we display the I/O and the CPU time of Dumpy, respectively.
For the rest rows, we show the timelines of building DumpyOS with different numbers of threads.
We sum the I/O or CPU time of each stage and show them in the figure.
Note that in Stages 1 and 4, the tasks are executed in batches, and in each batch, CPU and I/O threads need to synchronize on the inner barriers (see Figure~\ref{fig:DumpyOS-timeline}).
To make the figure clear, we omit the detailed timelines between inner barriers, and sum the total time consumed in these stages.
In the synchronization process of these two stages, CPU threads always need to wait for I/O threads.
We mark the (sum) waiting time with the grey arrows.
For Stage 1, CPU time can be totally masked by I/O time, while in Stage 4, a small portion of CPU time cannot be overlapped by I/O time; we illustrate this with a small non-overlapping part at the ends of the bars in the figure.

Using parallel computation, the CPU time costs for Stages 1 and 3 reduce almost linearly with the number of threads and can be fully masked by the simultaneous I/O-bound work.
Note that DumpyOS introduces one more pass of sequential read of the dataset in Stage 2, but this time can be overlapped by the CPU time.
Thus, this design improves the overall wall-clock time by reducing the writing time in Stage 4.
In Stage 4, our buffering technique significantly reduces the write time to \textbf{40\%} (see the comparison between Dumpy and 2-threads DumpyOS, where only one thread is used to issue I/O requests).
By parallelizing writing requests, the write time is further reduced by up to \textbf{2}x when using 5 threads.
Notice that with 5 threads, the CPU time is totally masked by the I/O time except for Stage 4, and the SSD is nearly saturated (where the write time is close to the sequential read time in Stage 4).
Only marginal benefits can be gained as the number of threads increases.
Finally, we note that more than \textbf{95\%} of the additionally introduced CPU time in Stage 4 is masked by the I/O time, verifying the effectiveness of our thread-scheduling algorithm.


\subsection{Query Processing}
In this section, we verify the accuracy of the query processing.
\begin{figure*}[tb]
\setlength{\abovecaptionskip}{-2pt}
\subfigcapskip=-1pt
\includegraphics[width=0.96\linewidth]{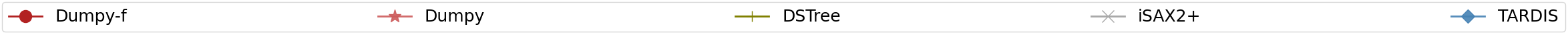} 
\centering
    \subfigure[Rand]{
		\centering
		\includegraphics[width=.2\linewidth]{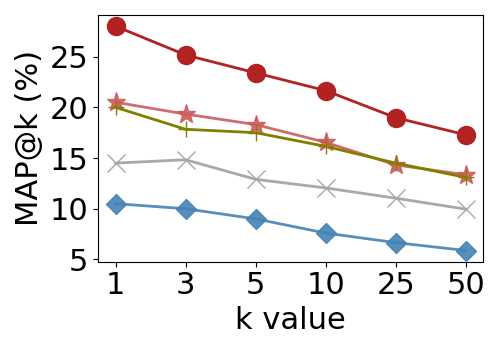}
		\label{fig:rand-recall}
	}
	\subfigure[DNA]{
		\centering	
		\includegraphics[width=.2\linewidth]{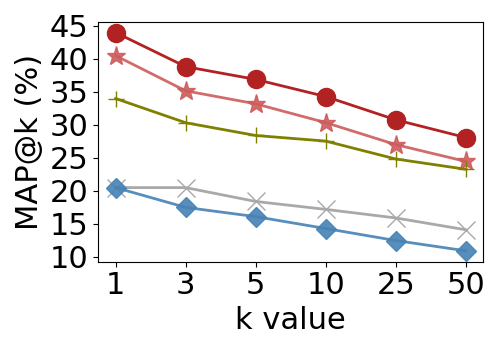}
		\label{fig:dna-recall}
	}
	\subfigure[ECG]{
		\centering
		\includegraphics[width=.2\linewidth]{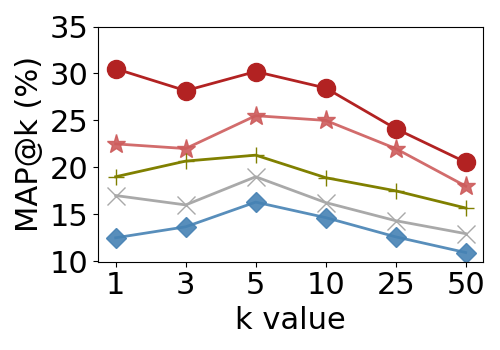}
		\label{fig:ecg-recall}
	}
	\subfigure[Deep]{
		\centering
		\includegraphics[width=.2\linewidth]{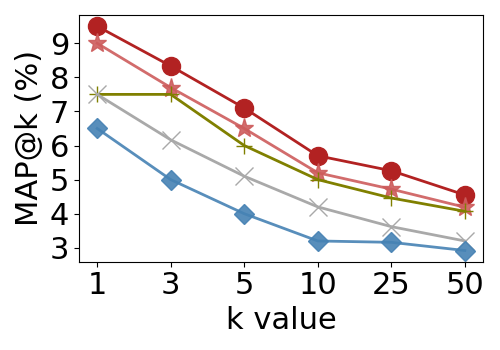}
		\label{fig:deep-recall}
	}
	
	\subfigure[Rand]{
		\centering
		\includegraphics[width=.2\linewidth]{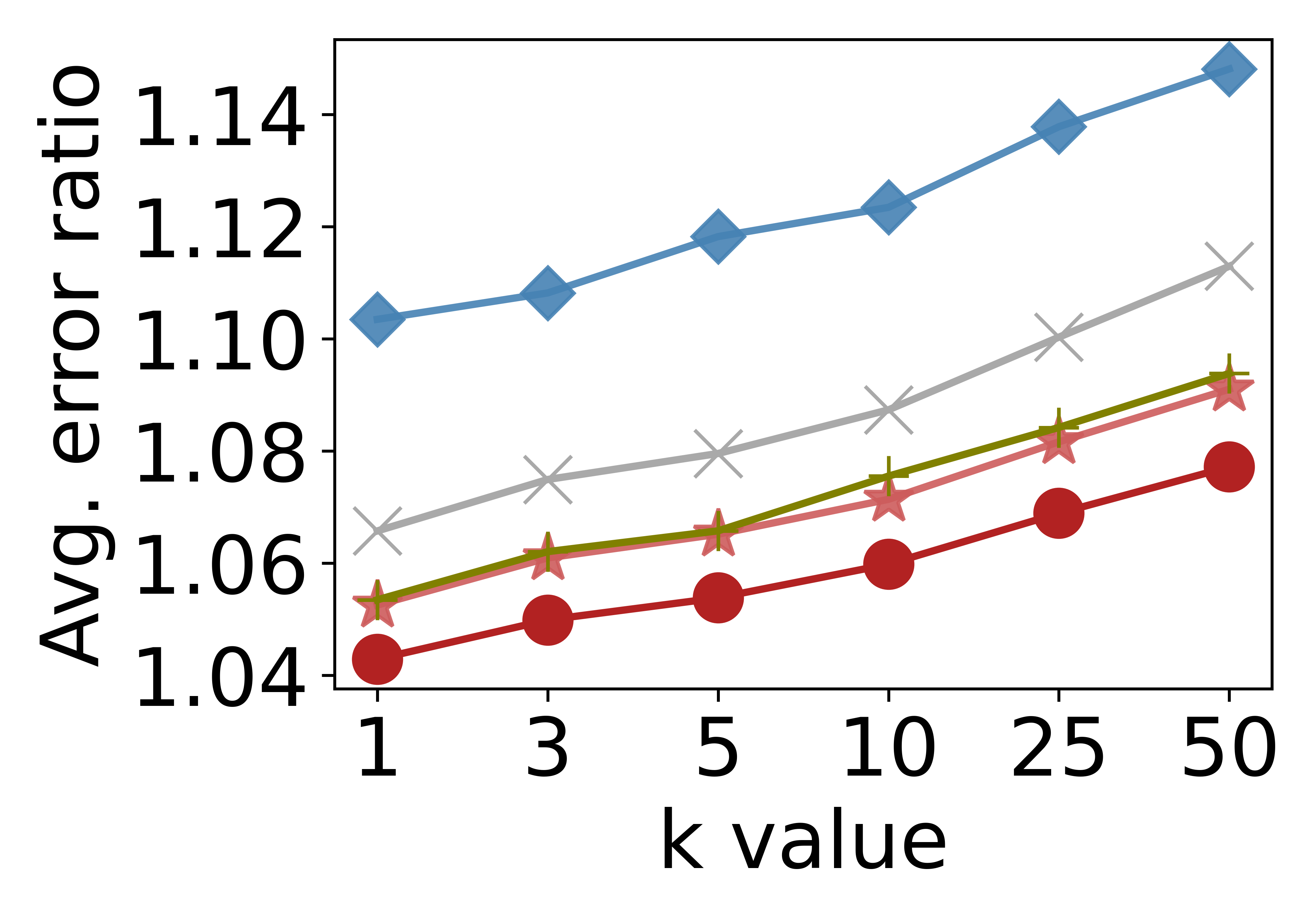}
		\label{fig:rand-precision}
	}
	\subfigure[DNA]{
		\centering	
		\includegraphics[width=.2\linewidth]{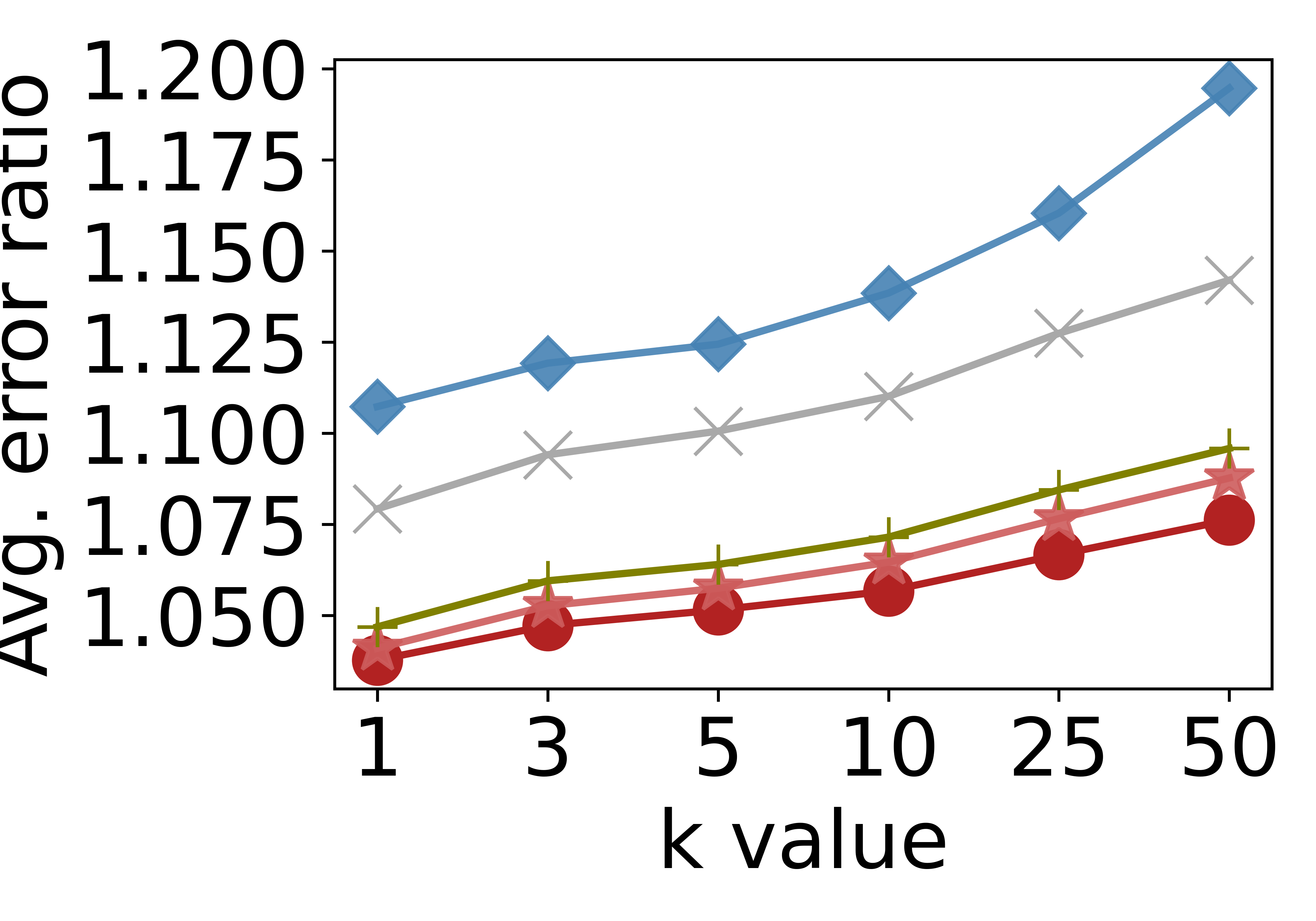}
		\label{fig:dna-precision}
	}
	\subfigure[ECG]{
		\centering
		\includegraphics[width=.2\linewidth]{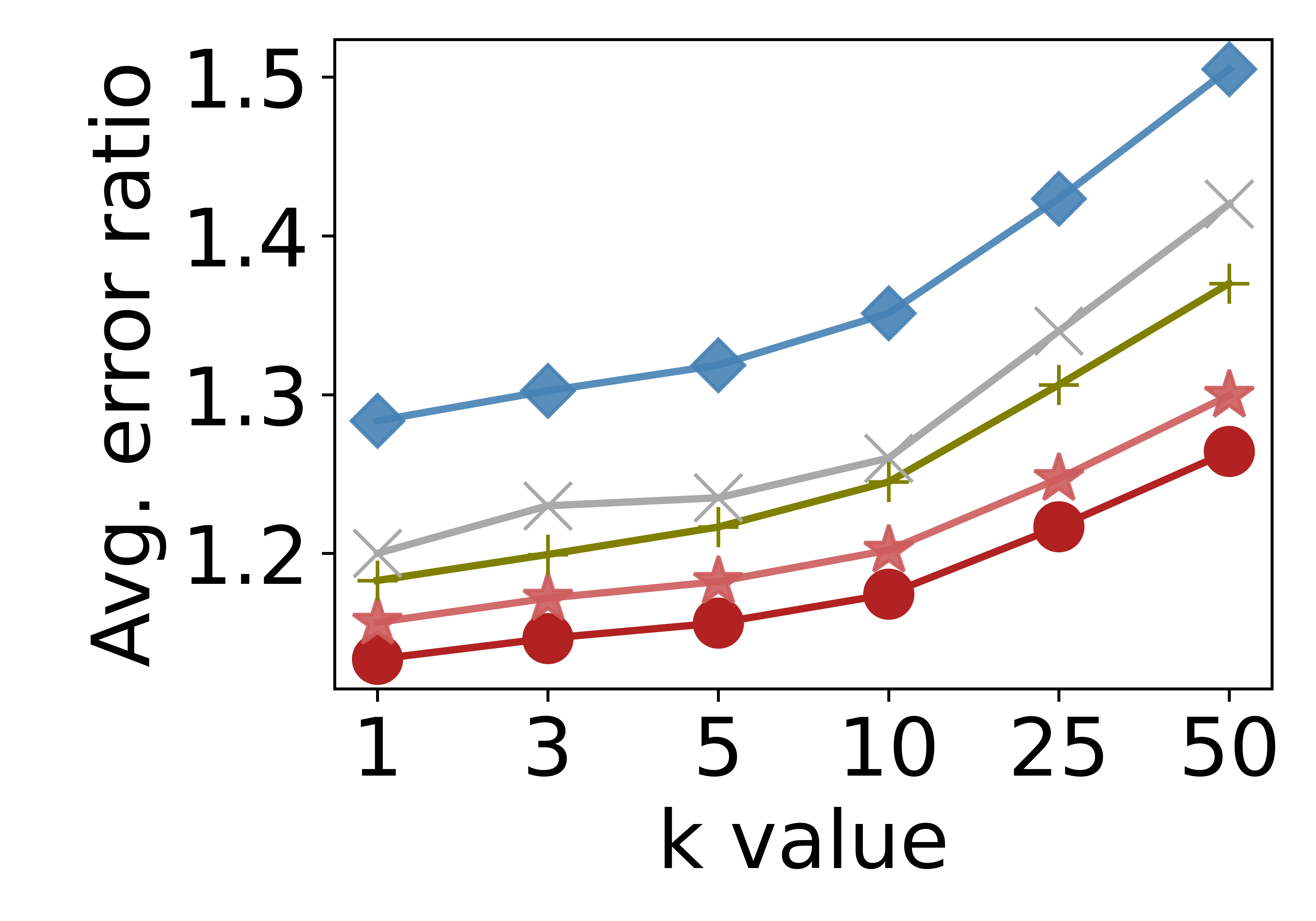}
		\label{fig:ecg-precision}
	}
	\subfigure[Deep]{
		\centering
		\includegraphics[width=.2\linewidth]{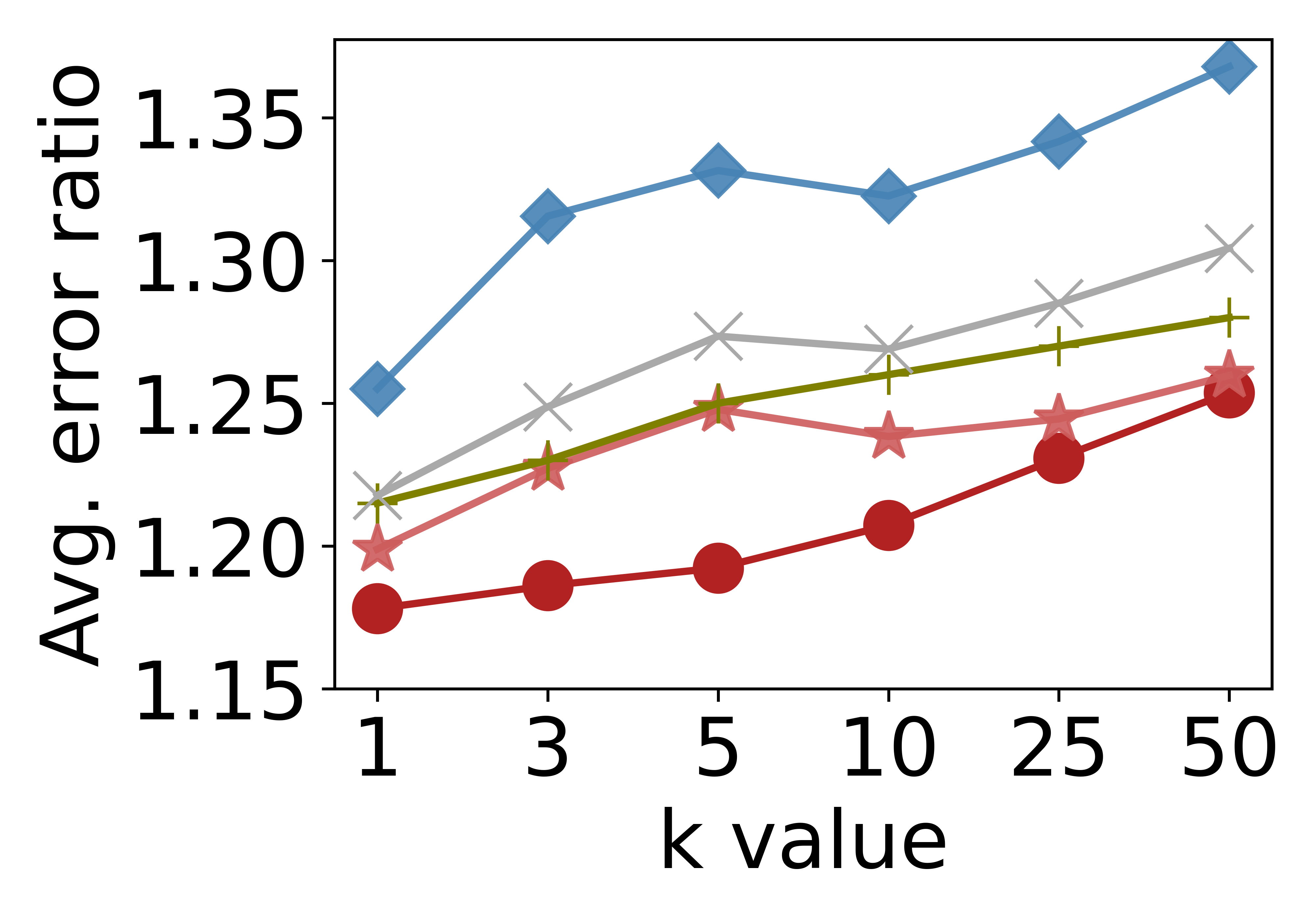}
		\label{fig:deep-precision}
	}
 \vspace{-3mm}
\caption{Approximate search under ED (search one node).}
\label{fig:approx} 
\vspace{-1mm}
\end{figure*}

\begin{figure*}[tb]
\setlength{\abovecaptionskip}{-2pt}
\subfigcapskip=-1pt
\centering
    \subfigure[Rand]{
		\centering
		\includegraphics[width=.2\linewidth]{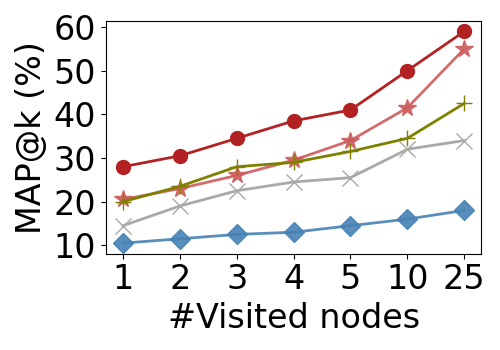}
		\label{fig:rand-node-recall}
	}
	\subfigure[DNA]{
		\centering	
		\includegraphics[width=.2\linewidth]{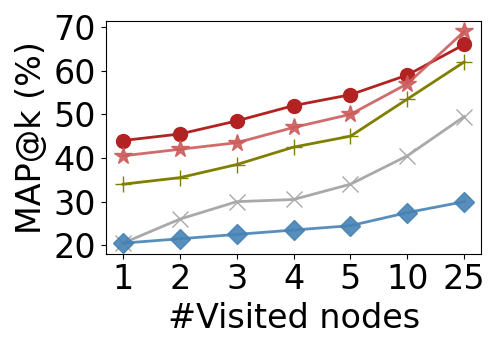}
		\label{fig:dna-node-recall}
	}
	\subfigure[ECG]{
		\centering
		\includegraphics[width=.2\linewidth]{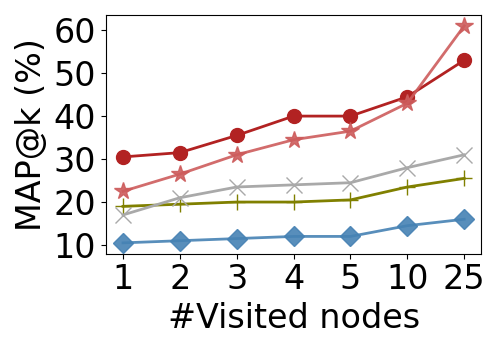}
		\label{fig:ecg-node-recall}
	}
	\subfigure[Deep]{
		\centering
		\includegraphics[width=.2\linewidth]{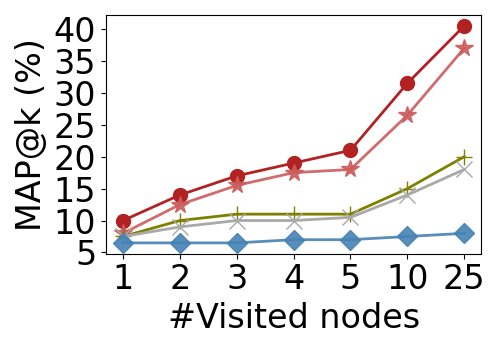}
		\label{fig:deep-node-recall}
	}
	
	\subfigure[Rand]{
		\centering
		\includegraphics[width=.2\linewidth]{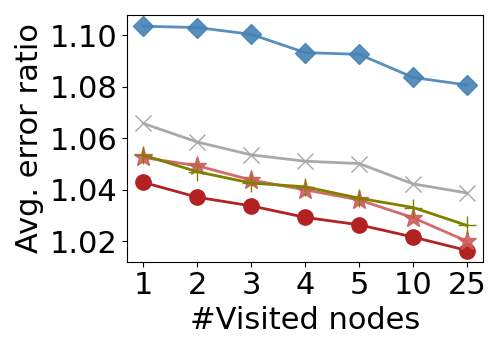}
		\label{fig:rand-node-precision}
	}
	\subfigure[DNA]{
		\centering	
		\includegraphics[width=.2\linewidth]{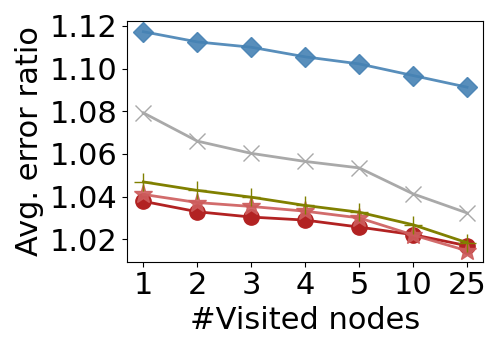}
		\label{fig:dna-node-precision}
	}
	\subfigure[ECG]{
		\centering
		\includegraphics[width=.2\linewidth]{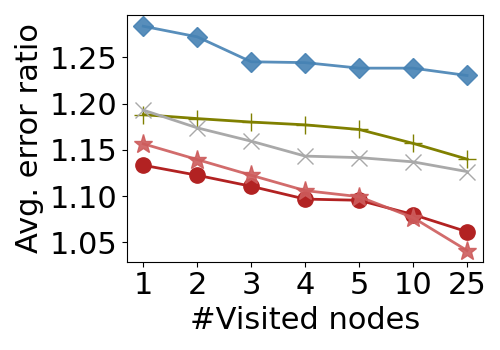}
		\label{fig:ecg-node-precision}
	}
	\subfigure[Deep]{
		\centering
		\includegraphics[width=.2\linewidth]{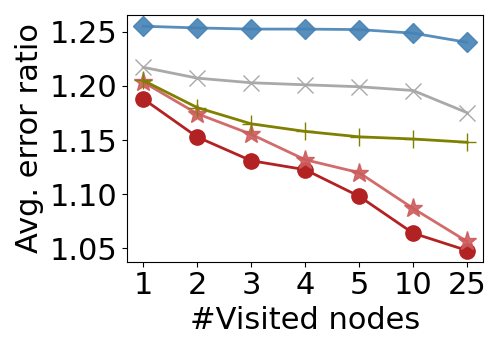}
		\label{fig:deep-node-precision}
	}
  \vspace{-3mm}
\caption{Extended approximate search under ED (k=1).}
\label{fig:approx-node} 
  \vspace{-1mm}
\end{figure*}

\begin{figure}[tb]
  \includegraphics[width=0.99\linewidth]{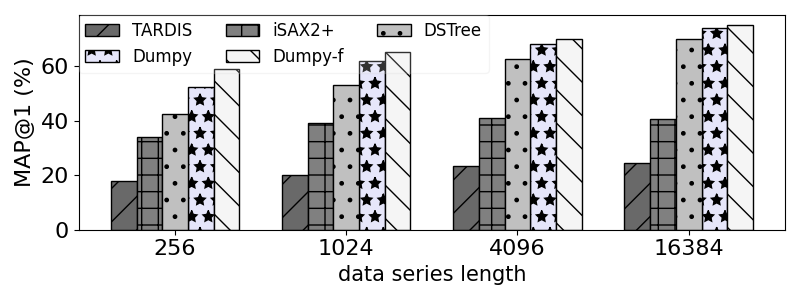}
  \caption{Approx. Search vs. series lengths (search 25 nodes)}
  \label{fig:app-length}
  \vspace{-6mm}
\end{figure}

\begin{figure}[tb]
\includegraphics[width=\linewidth]{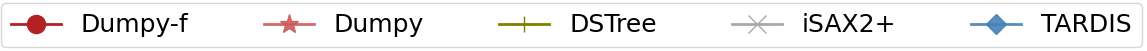}
\subfigure[$k=1$]{
  \includegraphics[width=0.47\linewidth]{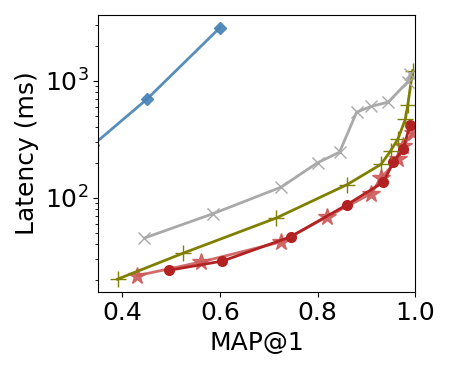}
  \label{fig:ng-1}
}
\subfigure[$k=50$]{
\includegraphics[width=0.47\linewidth]{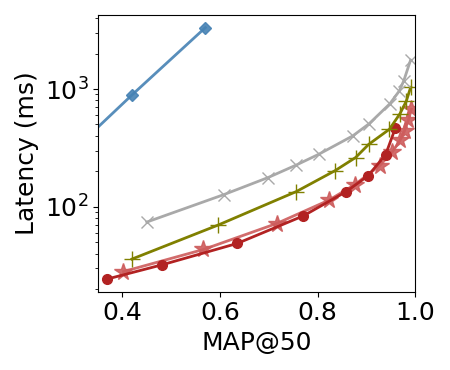}
\label{fig:ng-50}
}
\caption{Efficiency vs. accuracy.}
\label{fig:ng} 
\vspace{-6mm}
\end{figure}

\begin{figure*}[tb]
\centering
\includegraphics[width=0.9\linewidth]{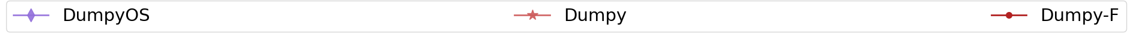}
\subfigure[Rand]{
  \includegraphics[width=0.2\linewidth]{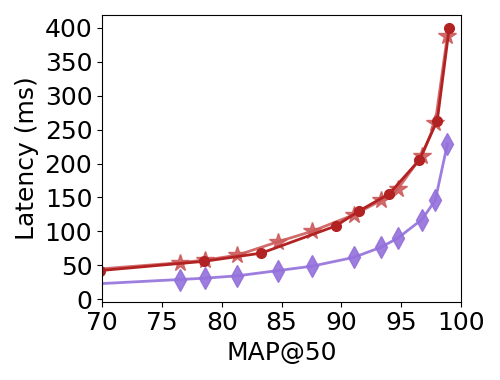}
}
\subfigure[DNA]{
  \includegraphics[width=0.2\linewidth]{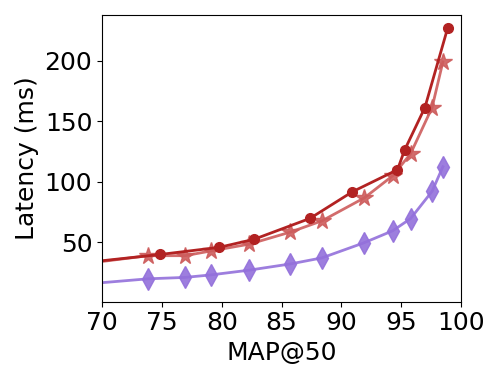}
}
\subfigure[ECG]{
  \includegraphics[width=0.2\linewidth]{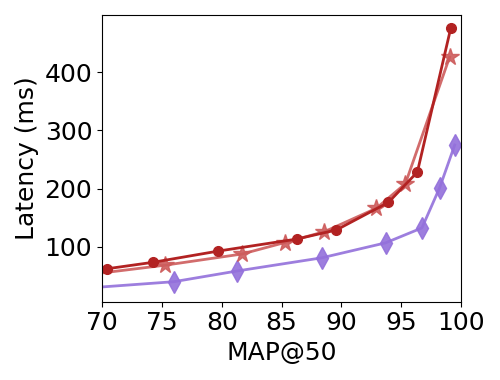}
}
\subfigure[Deep]{
  \includegraphics[width=0.2\linewidth]{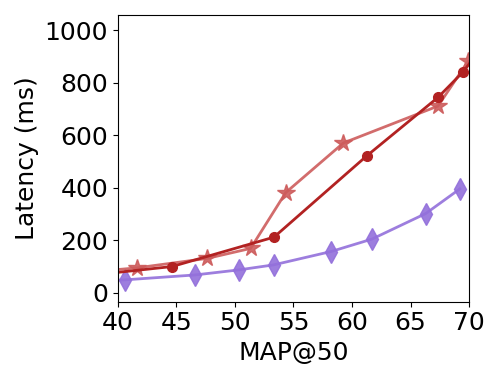}
}
\caption{Parallel pruning-based $ng$-approximate search with DumpyOS ($k$=50).}
\label{fig:ng-DumpyOS} 
\vspace{-3mm}
\end{figure*}

\begin{figure*}[tb]
\centering
\subfigure[Rand]{
  \includegraphics[width=0.2\linewidth]{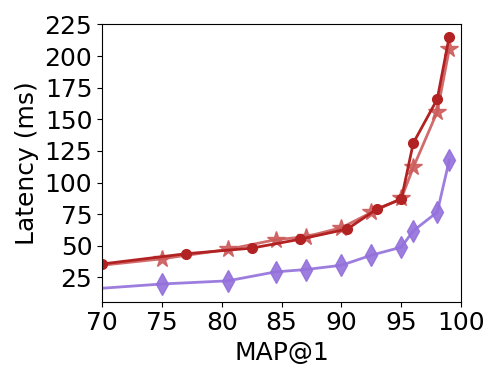}
}
\subfigure[DNA]{
  \includegraphics[width=0.2\linewidth]{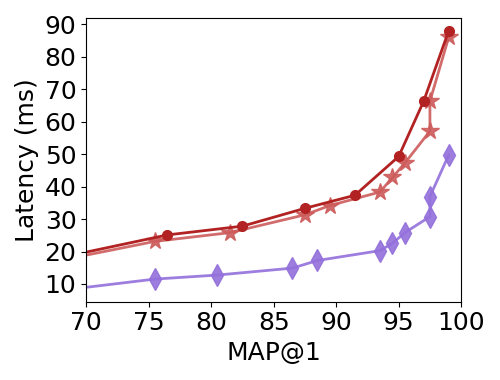}
}
\subfigure[ECG]{
  \includegraphics[width=0.2\linewidth]{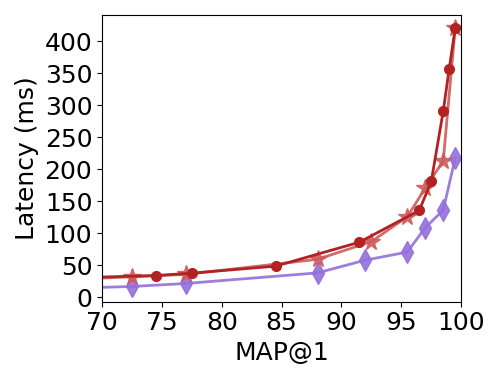}
}
\subfigure[Deep]{
  \includegraphics[width=0.2\linewidth]{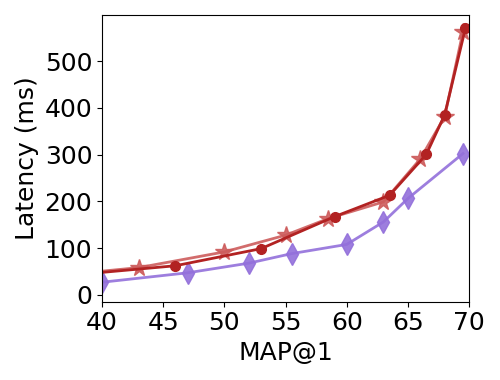}
}
\caption{Parallel pruning-based $ng$-approximate search with DumpyOS ($k$=1).}
\label{fig:ng-DumpyOS-k=1} 
\vspace{-4mm}
\end{figure*}

\begin{figure}[tb]
\includegraphics[width=\linewidth]{1-col-title-4.png}
\subfigure[MAP]{
  \includegraphics[width=0.46\linewidth]{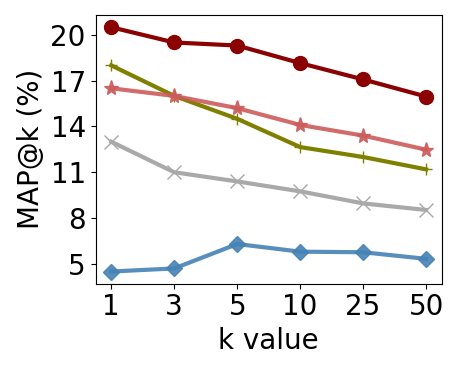}
  \label{fig:dtw-recall}}
\subfigure[Avg. error ratio]{
\includegraphics[width=0.46\linewidth]{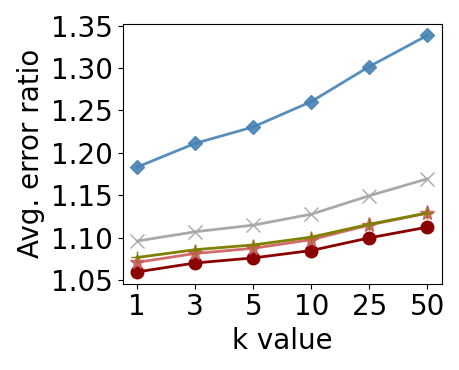}
\label{fig:dtw-er}
}
\caption{Approx. search under DTW (search one node)}
\label{fig:dtw} 
\vspace{-2mm}
\end{figure}

\begin{figure}[tb]
\includegraphics[width=\linewidth]{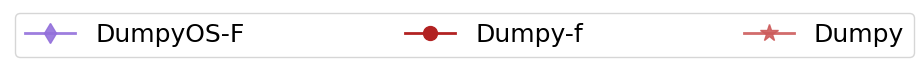}

\subfigure[Rand]{
  \includegraphics[width=0.47\linewidth]{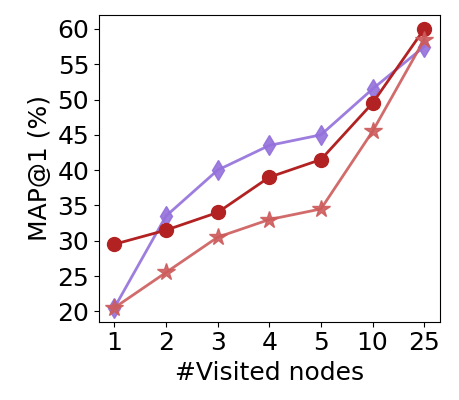}
  \label{fig:dumpy-vf-rand-1}
}
\subfigure[DNA]{
\includegraphics[width=0.47\linewidth]{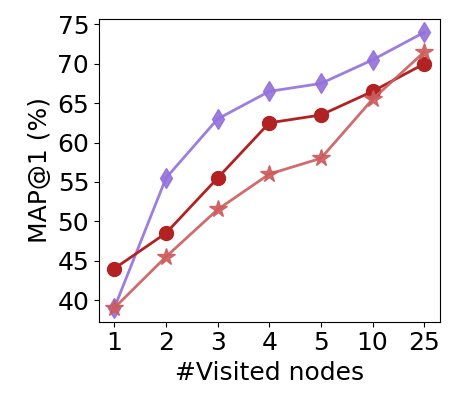}
\label{fig:dumpy-vf-dna-1}
}

\subfigure[Rand]{
  \includegraphics[width=0.47\linewidth]{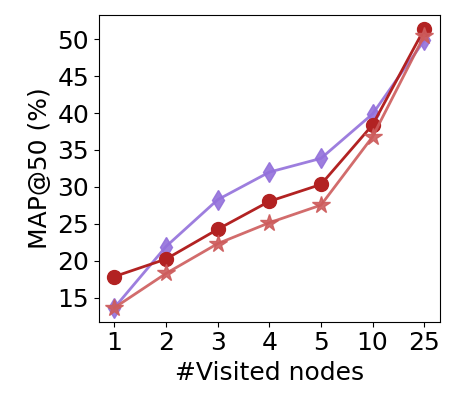}
  \label{fig:dumpy-vf-rand-50}
}
\subfigure[DNA]{
\includegraphics[width=0.47\linewidth]{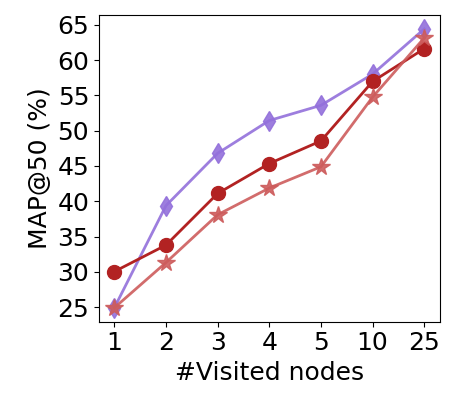}
\label{fig:dumpy-vf-dna-50}
}
\caption{Extended approximate search with DumpyOS-F.}
\label{fig:dumpy-vf} 
\vspace{-3mm}
\end{figure}

\subsubsection{Approximate Search}
\label{sec:expr-approx-search}
First, we evaluate the accuracy of approximate results across four datasets. 

\noindent\textbf{[Search one node]} 
We compare these approaches when searching one node to obtain the approximate top-$k$ result on three datasets with ED distance, and results are shown in Figure~\ref{fig:approx}. It can be seen that Dumpy consistently outperforms other approaches.
Specifically, Dumpy improves the average accuracy by \textbf{84}\%, \textbf{46}\%, \textbf{11}\% and reduces the average error ratio by \textbf{7.3}\%, \textbf{3.4}\% and  \textbf{1.4}\% on the four datasets compared with TARDIS, iSAX2+ and DSTree respectively. Tardis has the lowest performance, due to its low fill factor. 
iSAX2+ suffers from insufficient intra-node series proximity, characterized by the number of uneven nodes ($>$\textbf{20}\% leaf nodes have one segment using more than 4 bits than other segments).
This number is only \textbf{1.4}\% for Dumpy.
As for TARDIS, the small-sized leaves cannot provide enough candidate series, though the series in the target leaf have a superior quality (small distance to the query). 
Moreover, Dumpy-Fuzzy has higher accuracy than Dumpy and other approaches, which verifies our duplication strategy.

\noindent\textbf{[Search multiple nodes]} 
In Figure~\ref{fig:approx-node}, we compare the accuracy of searching multiple nodes (1 to 25) for top-1 result with ED distance.
The MAP value of Dumpy and Dumpy-$f$ increases remarkably faster than the competitors, attributed to our multi-ary structure that provides closer sibling nodes.
When visiting 25 nodes, Dumpy and Dumpy-$f$ improve the accuracy by \textbf{58}\%, \textbf{65}\% and reduce the average error ratio by \textbf{3.6}\% and \textbf{3.7}\% on average of four datasets respectively compared with the second-best approach, DSTree.
We also compare the accuracy as the series length varies (Figure~\ref{fig:app-length}). The accuracy on different lengths shares similar rankings.

\noindent\textbf{[Efficiency vs accuracy]} 
We extend the approximate search to all leaf nodes with lower bound pruning to evaluate the indexes' response time under the whole MAP range.
The results are depicted in Figure~\ref{fig:ng}.
Benefiting from the high-proximity nodes and compact index structure,
Dumpy surpasses its competitors in both low and high-precision intervals.
Our results show that Dumpy can achieve 60\%-70\% MAP within 100ms on TB-level datasets, while its index building time is 4x faster than the SOTA competitors. 
These results demonstrate that Dumpy fulfills the requirements of many kNN-based applications.

\noindent\textbf{[Parallel search of DumpyOS]} 
In Figures~\ref{fig:ng-DumpyOS} and ~\ref{fig:ng-DumpyOS-k=1}, we show the $ng$-approximate query performance improvements of DumpyOS against Dumpy and Dumpy-$f$. 
The queries are processed sequentially.
DumpyOS accelerates the pruning-based query process with the parallel search algorithm in Section~\ref{sec:DumpyOS-query}.
DumpyOS reduces the query latency of Dumpy by \textbf{41}\%, \textbf{44}\%, \textbf{36}\%, \textbf{55}\% on Rand, DNA, ECG and Deep datasets, respectively when $k$=50 on the high-recall range.
When $k$=1, DumpyOS reduces the query latency of Dumpy by \textbf{43}\%, \textbf{43}\%, \textbf{35}\%, and \textbf{46}\% on Rand, DNA, ECG, and Deep datasets, respectively on the high-recall range.
The speedup of querying becomes larger as MAP goes higher since a higher precision requires more nodes to be loaded and more distance calculations, which can be accelerated by the buffering technique and the parallel computation of DumpyOS.

\noindent\textbf{[Searching under DTW]} 
In this experiment, we compare the accuracy under DTW distance in Figure~\ref{fig:dtw}.
Due to the inherent hardness, the precision is lower than ED generally.
However, Dumpy and Dumpy-$f$ still achieve better precision and error ratio under DTW.
Since the absolute distance of DTW is smaller than ED, the differences of the error ratio among all the methods tend to be smaller (except for TARDIS).

\subsubsection{Approximate search on DumpyOS-F}
We evaluate our novel, extended approximate search algorithm, DumpyOS-F, on Rand and DNA datasets with Dumpy and Dumpy-Fuzzy, as shown in Figure~\ref{fig:dumpy-vf}.
When visiting a single leaf, DumpyOS-F shares the same accuracy with Dumpy, because of the invariance of the index structure.
Dumpy-Fuzzy is the most accurate benefiting from the duplicated series stored in the node.
When we visit two or more nodes, DumpyOS achieves the highest accuracy, that is, an average (over our four settings) of \textbf{18}\% and \textbf{8.7}\%, respectively, higher MAP than Dumpy and Dumpy-Fuzzy.

\subsubsection{Exact Search}

We evaluate the exact search efficiency of Dumpy against other methods in Table~\ref{tab:exact}.
Since TARDIS does not support exact $k$NN search in the original paper, we implement a similar algorithm as the iSAX-index family, where the nodes summarized with iSAX words can be pruned during searching.
The results are reported on average of 40 queries with $k$=50.
Overall, Dumpy achieves the best efficiency in all cases. It is worth noting that although DSTree has a higher pruning ratio than Dumpy, the response time is still slower than Dumpy. The reason is as follows. DSTree takes a longer time to compute the lower bound of distance due to computing the standard deviation frequently. iSAX2+ suffers from the low fill factor and needs to read about \textbf{3} times nodes more than Dumpy and DSTree.

\begin{table}[bt]
{
\scriptsize
	\centering
	\caption{Exact Search statistics. The response time is also broken down into I/O time (the first number in the parentheses) and CPU time (the second number).}
	\label{tab:exact}
	\begin{tabular}{c|c|c|c|c}
		\hline
		    & Method & Resp. time (s) &\thead{\#Loaded\\ Nodes}  & \thead{Prune\\ ratio} \\
		\hline\hline
		\multirow{4}{*}{\thead{Rand\\-ED}} & iSAX2+ & 65 (50+15) & 7595   &81.51\% \\
		\cline{2-5}
		&                 DSTree & 33 (20+13) & 2027   & 86.06\%\\
		\cline{2-5}
		&                TARDIS   & 53 (15+38) & 1665744 & 59.30\%\\
		\cline{2-5}
		& \textbf{Dumpy}   & \textbf{17} (\textbf{12}+\textbf{5}) & 2641 & 83.70\%\\
		\hline\hline
		\multirow{4}{*}{\thead{Rand\\-DTW}}  &iSAX2+ & 151 (85+66) & 18660   & 72.58\% \\
		\cline{2-5}
		&         DSTree & 79 (24+65) & 3678  & 75.12\%\\
		\cline{2-5}
		&         TARDIS   & 208 (22+186) & 2846868  &  49.24\% \\
		\cline{2-5}
		&         \textbf{Dumpy}   & \textbf{58} (\textbf{18}+\textbf{40}) & 3997  & 73.61\% \\
		\hline\hline
		\multirow{4}{*}{\thead{DNA\\-ED}} & iSAX2+ & 42 (26+16) & 1077   &91.04\% \\
		\cline{2-5}
		&     \textbf{DSTree} & 21 (16+5) & 326   & 94.00\%\\
		\cline{2-5}
		&          TARDIS   & 40 (11+29) & 161959 & 71.64\%\\
		\cline{2-5}
		& \textbf{Dumpy}   & \textbf{12} (\textbf{10}+\textbf{2}) & 433 & 91.69\%\\
		\hline\hline
		\multirow{4}{*}{\thead{DNA\\-DTW}}  & iSAX2+ & 116 (60+56)& 2163 & 87.77\% \\
		\cline{2-5}
		&         DSTree & 63 (18+45) & 497  & 90.93\%\\
		\cline{2-5}
		&         TARDIS   & 143 (16+127) & 194645 & 68.78\% \\
		\cline{2-5}
		&         \textbf{Dumpy}   & \textbf{53} (\textbf{14}+\textbf{39}) & 528 & 89.41\% \\
		\hline\hline
	\end{tabular}
    } 
\end{table}

\begin{figure}[tb]
\subfigure[Rand-ED]{
  \includegraphics[width=0.47\linewidth]{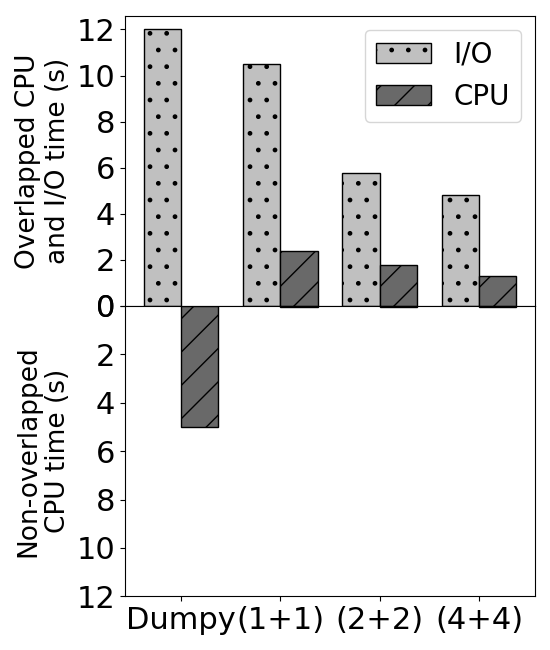}
  \label{fig:p-search-rand-ed}
}
\subfigure[DNA-ED]{
\includegraphics[width=0.47\linewidth]{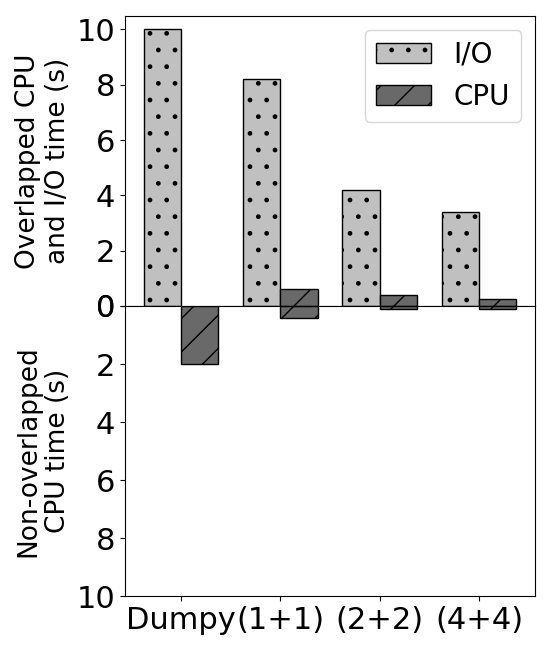}
\label{fig:p-search-dna-ed}
}

\subfigure[Rand-DTW]{
  \includegraphics[width=0.47\linewidth]{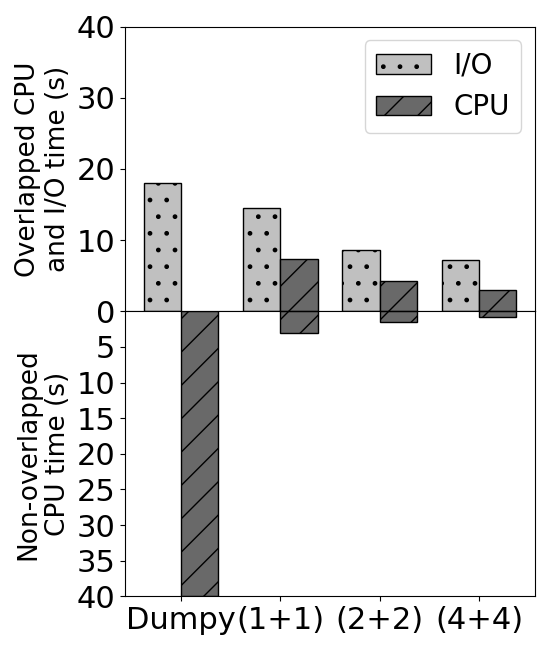}
  \label{fig:p-search-rand-dtw}
}
\subfigure[DNA-DTW]{
\includegraphics[width=0.47\linewidth]{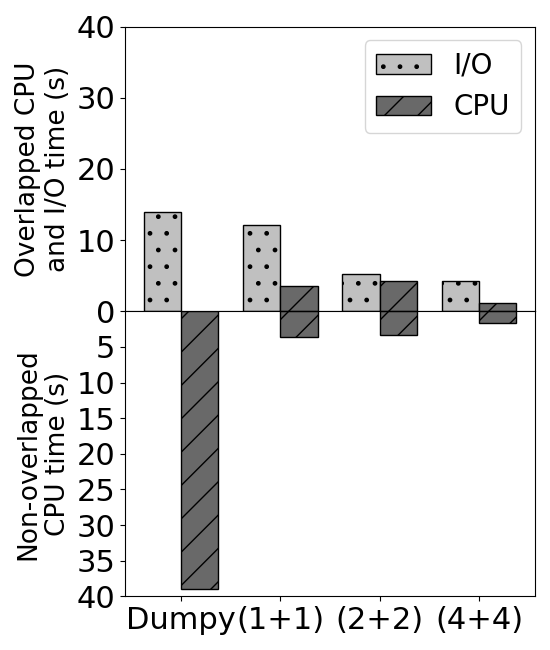}
\label{fig:p-search-dna-dtw}
}
\caption{Parallel exact search time with DumpyOS. The first column is for single-thread Dumpy and the rest three columns are for DumpyOS. On x-axis, $A$ + $B$ indicates $A$ threads for reading, and $B$ threads for computing.}
\label{fig:p-search} 
\end{figure}

\begin{figure}[tb]
\subfigure[I/O throughput monitoring during search.]{
  \includegraphics[width=0.47\linewidth]{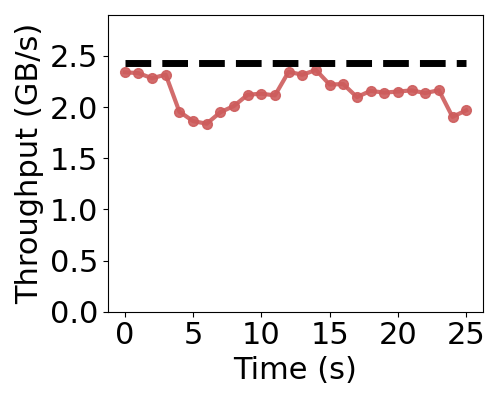}
  \label{fig:bandwidth}
}
\subfigure[The increase of I/O throughput with parallelization.]{
\includegraphics[width=0.47\linewidth]{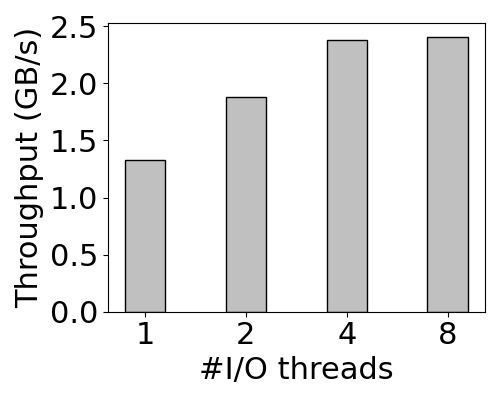}
\label{fig:bandwidth-inc}
}
\caption{I/O throughput on parallel exact search with DumpyOS.}
\label{fig:io-bottleneck} 
\vspace{-4mm}
\end{figure}

\noindent\textbf{[Multi-thread exact search with DumpyOS]}
In Figure~\ref{fig:p-search}, we show the exact query performance of DumpyOS when varying the number of threads.
To gain more insights, we show the I/O and CPU time of the query time components of the total query time, as well.
I/O time is shown in the upper part.
CPU time is broken into two parts.
One part overlaps with the I/O time, shown in the upper part.
The other, non-overlapped part, is shown in the lower part.

As a single-thread algorithm, Dumpy's I/O and CPU time are not overlapped at all.
In contrast, DumpyOS allows I/O threads and CPU threads to work simultaneously, therefore, their execution times are overlapping.
The overall query time is computed by the sum of I/O time (in the upper part) and non-overlapping CPU time (in the bottom part).

Compared with Dumpy, DumpyOS achieves \textbf{5.8x} faster query time on average with "4+4" threads, where I/O time is \textbf{2.8x} faster while CPU time is \textbf{7.2x} faster.
The acceleration is obvious from one to two I/O threads (see DumpyOS "1+1" to "2+2"), which comes from our buffering and parallel reading algorithms.
As we keep increasing the number of threads, the SSD's throughput turns saturated, and the performance reaches a peak.
The SIMD technique significantly reduces computing time, especially under DTW distance (see Dumpy to DumpyOS "1+1").
Observe that the CPU time is almost totally masked by the I/O time (see DumpyOS "4+4"), as expected.

\noindent\textbf{[Hardware bottleneck]}
Since in the pruning-based search process of DumpyOS, the I/O time becomes the major part of the total wall-clock time, we further study the bottleneck of the hardware that limits the I/O performance.
Given that nearly all the I/O requests are sequential I/Os, we measure the I/O performance with the (read) throughput.
The results are shown in Figure~\ref{fig:io-bottleneck}.
In Figure~\ref{fig:bandwidth}, we monitor the I/O throughput during a period of the pruning-based parallel exact search with 4 threads on the Rand100GB dataset.
It can be seen that the I/O throughput is very close to the I/O bandwidth, which is measured using the fio~\cite{fio} command under the condition of multi-threads sequential read.
This indicates that the bandwidth of the NVMe SSD is nearly saturated, especially when the query needs to load a large number of nodes.

In Figure~\ref{fig:bandwidth-inc}, we compare the I/O throughput with a different number of I/O threads.
Before 4 threads, the read throughput increases as more threads are used to submit I/O requests.
This is because more threads produce a number of I/O requests that lead to a better use of the parallelization ability of the NVMe SSD.
However, with more than 4 I/O threads, the SSD bandwidth becomes saturated, which limits further improvement in query performance.
This indicates that under a better hardware environment, including a more advanced PCIe channel (e.g., version of $\geq$ 4.0), a more efficient I/O system (e.g., SPDK~\cite{spdk}) or a better SSD, DumpyOS can further improve its I/O time, and thus the overall query performance.

\begin{figure}[tb]
\centering
  \includegraphics[width=0.7\linewidth]{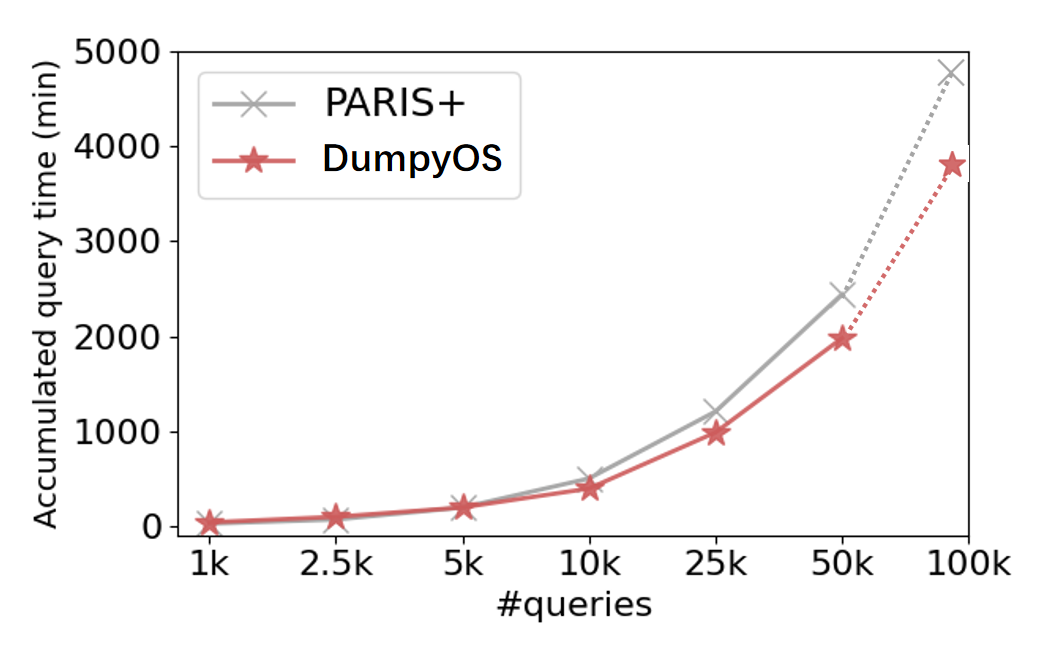}
  \caption{DumpyOS v.s. PARIS+.}
  \label{fig:paris}
  \vspace{-2mm}
\end{figure}

\noindent\textbf{[Pruning-based search vs. scan-based search]}
We further compare DumpyOS with SOTA parallel-scan-based exact search algorithm PARIS+, and report the results in Figure~\ref{fig:paris}.
During the search process, PARIS+ needs to materialize the leaves that have been visited, by flushing these data when the memory buffer is full.
In our experiments, PARIS+ shows slight superiority when the number of served queries is small.
In this case, the buffer is not full, hence no disk writing.
Yet, as more queries come, DumpyOS becomes more efficient since the random disk writes hurt the performance of PARIS+.
Even after serving 50,000 queries, the ratio of full leaves in PARIS+ is only \textbf{16\%}, at which time PARIS+ has generated over 300,000 leaves.
On the contrary, DumpyOS that leverages parallel techniques to achieve faster index building, leaves no indexing burden on the query stage and hence achieves better query performance with intensive query workloads.



\subsection{Complete Workloads}
\begin{figure}[tb]
  \includegraphics[width=0.9\linewidth]{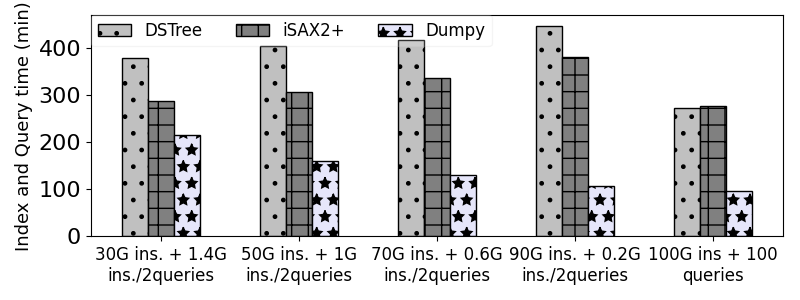}
  \caption{Update performance (4GB memory)}
  \label{fig:conmplte-workload}
  \vspace{-1mm}
\end{figure}
Finally, we compare different approaches when inserting new data series (Figure~\ref{fig:conmplte-workload}).
We omit TARDIS since it is designed for the static dataset and not easy to be extended.
To be fair, we implement all methods using a single thread (even though Dumpy is multi-threaded).
We use different synthetic workloads consisting of 100 exact queries, and a total of 100 million series, where queries are interleaved by a batch of insertions.
The results show Dumpy outperforms the competitors for all workloads, thanks to its compact structure.
Although the re-splitting and re-packing procedures add an additional cost, this cost is balanced by the efficiency improvements that these two designs bring along.
Moreover, Dumpy shows better performance when the initial batch size increases (while iSAX and DSTree show worse performance), because fewer insertions incur fewer re-splitting and re-packing actions.

\begin{figure}[tb]
\begin{minipage}{.47\linewidth}
    \includegraphics[width=\linewidth]{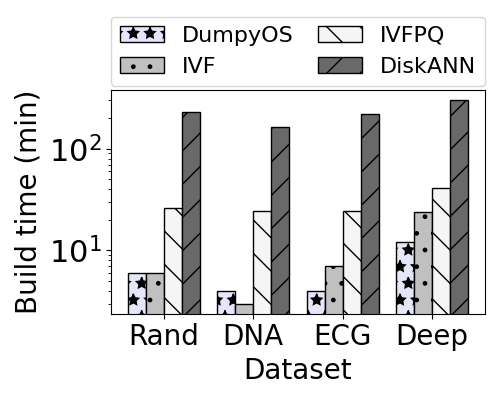}
    \caption{Index construction time on four 25GB datasets.}
    \label{fig:ann-build}
\end{minipage}%
\hspace{3mm}
\begin{minipage}{.47\linewidth}
    \includegraphics[width=\linewidth]{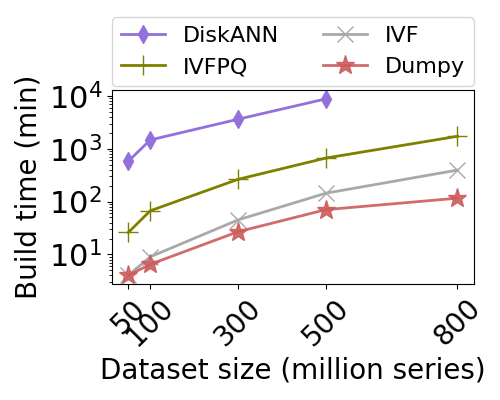}
    \caption{Index construction on Rand datasets.}
    \label{fig:diskann-build}
\end{minipage}%
\vspace{-2mm}
\end{figure}


\begin{figure*}[tb]
\setlength{\abovecaptionskip}{-4pt}
\subfigcapskip=-1pt
\centering
\includegraphics[width=0.9\linewidth]{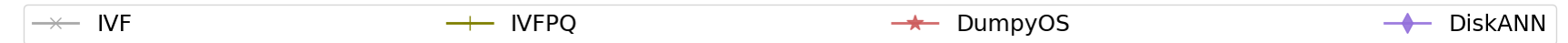}
\subfigure[Rand25GB]{
\centering
  \includegraphics[width=0.2\linewidth]{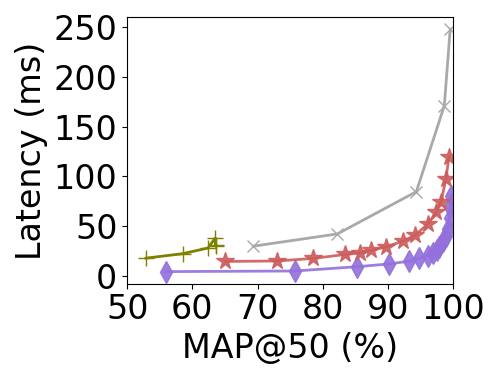}
}
\subfigure[DNA25GB]{
  \includegraphics[width=0.2\linewidth]{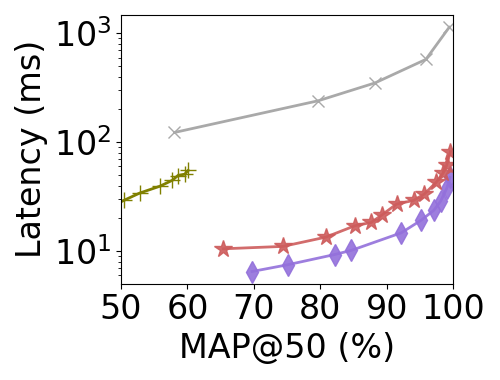}
}
\subfigure[ECG25GB]{
  \includegraphics[width=0.2\linewidth]{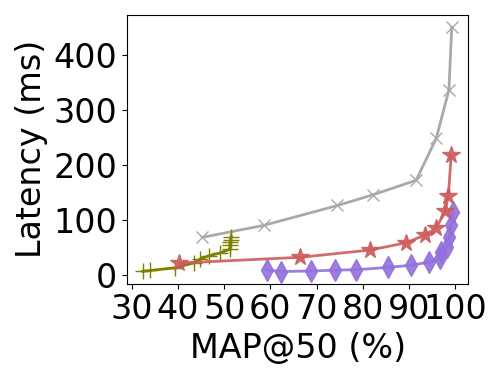}
}
\subfigure[Deep25GB]{
  \includegraphics[width=0.2\linewidth]{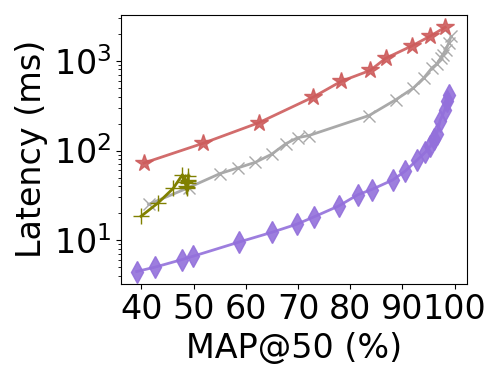}
}
\vspace{-2mm}
\caption{Query performance comparisons with high-dimensional vector indexes ($k$=50).}
\label{fig:ann-recall} 
\end{figure*}

\begin{figure*}[tb]
\setlength{\abovecaptionskip}{-4pt}
\subfigcapskip=-1pt
\centering
\includegraphics[width=0.9\linewidth]{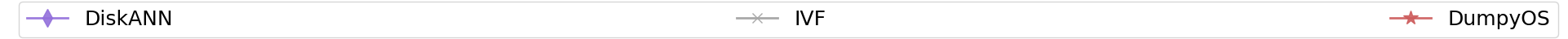}
\subfigure[Rand25GB]{
  \includegraphics[width=0.2\linewidth]{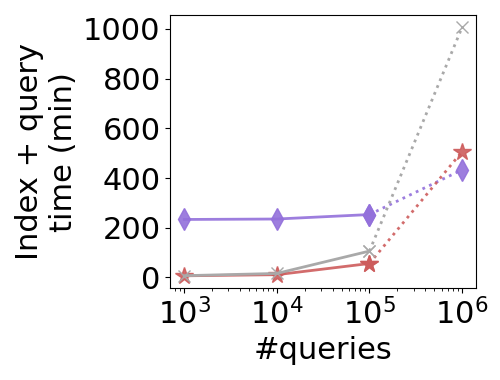}
}
\subfigure[DNA25GB]{
  \includegraphics[width=0.2\linewidth]{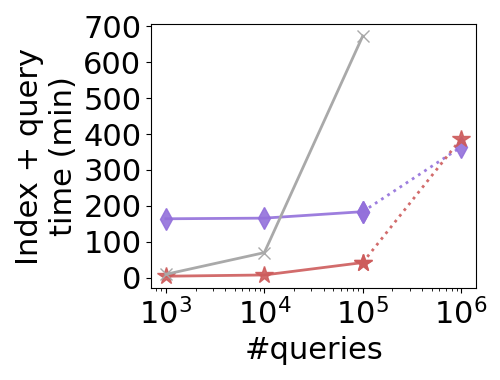}
}
\subfigure[ECG25GB]{
  \includegraphics[width=0.2\linewidth]{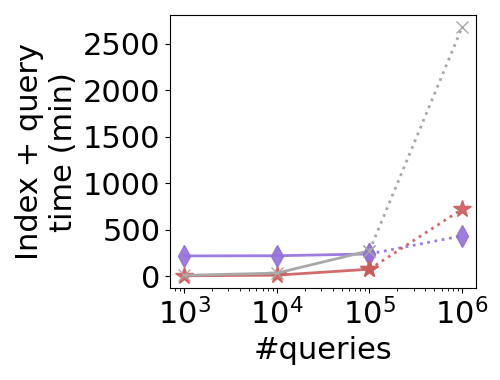}
}
\subfigure[Deep25GB]{
  \includegraphics[width=0.2\linewidth]{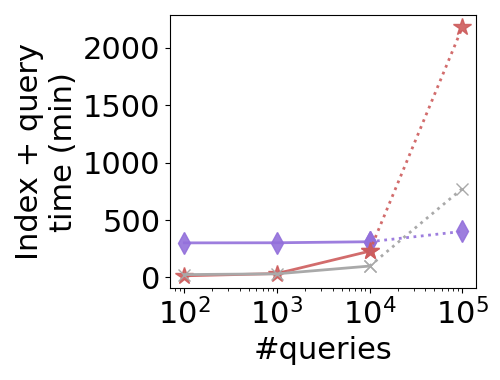}
}
\vspace{-2mm}
\caption{Accumulated time of indexing and $ng$-approximate queries under 90\% recall@50.}
\label{fig:accum} 
\vspace{-3mm}
\end{figure*}

\subsection{Comparison with High-d Vector Indexes}

In this subsection, we compare our DumpyOS solutions with SOTA high-dimensional vector indexes w.r.t. the building efficiency and $ng$-approximate query performance.
Note that SOTA high-dimensional vector indexes do not support neither exact search nor approximate search with quality guarantees.
Therefore, these approaches target a different set of applications than DumpyOS. Nevertheless, we compare to them for completeness.
\subsubsection{Experimental settings}
\noindent\textbf{[Algorithms and implementations.]}
We use \textbf{Disk-}
\noindent\textbf{ANN}~\cite{diskann} as a representative disk-based graph-based index and adopt the most recent implementation~\footnote{https://github.com/microsoft/DiskANN}.
We use \textbf{IVF}~\cite{faiss} and \textbf{IVFPQ}~\cite{pq} from the family of partition-based indexes.
We train the centroids of both, as well as the codebooks of IVFPQ with the Faiss library~\footnote{https://github.com/facebookresearch/faiss.git}.
Since there are no disk-based implementations of IVF and IVFPQ, we provide our own implementations; we store sequentially the data, as well as the quantized data of each cluster.
In this way, data in each cluster can be read with sequential I/Os when querying, similarly to DumpyOS.

\noindent\textbf{[Parameters.]}
For DiskANN, we set $R$=32 and $L$=100 for constructing the graph.
For IVF and IVFPQ, we randomly select 1 million data series from the datasets to train the centroids, and the number of centroids is set to be 1/10,000 of the number of the series.
For IVFPQ, the number of segments is set to 16 like DumpyOS.
The memory constraints when building indexes and running queries are controlled to be the same as Dumpy.
When building the index, all 10 threads are utilized. 

\noindent\textbf{[Datasets.]}
Since building DiskANN on 100GB datasets costs over one day which significantly exceeds the time limit of common data series similarity search applications, we extract subsets of the datasets of 25GB for the experiments in this subsection.

\subsubsection{Experimental results}
\noindent\textbf{[Index building.]}
In Figure~\ref{fig:ann-build}, we show the index-building time of the four indexes.
DumpyOS and IVF are the most efficient indexes, followed by IVFPQ, and finally, DiskANN.
Compared with IVFPQ and DiskANN, DumpyOS reduces the building time by 79\% and 97\%, respectively.
Note that the index-building complexity of IVF and IVFPQ is highly sensitive to the number of centroids.
In contrast, DumpyOS builds the index based on the iSAX summarization without any real distance calculations and thus achieves superior efficiency and robustness.
In Figure~\ref{fig:diskann-build}, we observe that DumpyOS has the best scalability, followed by IVF, IVFPQ, and DiskANN.
Note that on the Rand500GB dataset, DiskANN requires over 6 days to build the index, which is impractical for many scenarios.

\noindent\textbf{[Query performance.]}
In Figure~\ref{fig:ann-recall}, we compare the $ng$-approximate query performance of Dumpy with the other three indexes. 
We observe that DiskANN exhibited the best performance, followed by DumpyOS, IVF and IVFPQ.
Note that only the quantized codes are stored in the IVFPQ index, and this information loss of the quantization prevents IVFPQ from achieving a high recall.
DiskANN is slightly better than DumpyOS on data series datasets, while IVF and IVFPQ are much slower than DumpyOS.
In the classical long data series dataset like DNA, IVF suffers from the curse of dimensionality~\cite{dist-concen} and thus a poor clustering quality, which results in performance degradation.
On the other hand, DumpyOS can capture the differences on different segments between different data series and thus provide a superior query performance.
Note that a prominent characteristic of these data series is that they contain temporal semantics, i.e., the value continuity among the adjacent time axis.
Therefore, data series indexes that use summarization techniques such as PAA and SAX work well on these datasets.
On the contrary, on the Deep dataset, whose series exhibit high frequencies, DumpyOS is less effective in producing the series summarizations, and performs worse than IVF.

In Figure~\ref{fig:accum}, we show the accumulated time of indexing and $ng$-approximate query answering for 90\% recall@50.
Note that IVFPQ is skipped in this experiment since it cannot reach 90\% recall@50.
We extract 100,000 queries from the base 100GB datasets for this experiment, and we extrapolate the time on 1 million queries based on the average time of the 100,000 queries (i.e. the dotted lines).
As we observe, on Rand25GB, DNA25GB, and ECG25GB, only after running nearly 1 million queries, DiskANN is faster than DumpyOS.
For Deep25GB, DiskANN still needs 100,000 queries to compensate for the long building time.

\begin{figure}[tb]
\centering
  \includegraphics[width=0.7\linewidth]{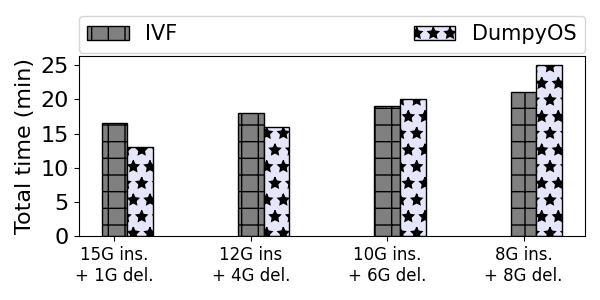}
  \caption{Update performance: the overall time of building a static index on 10GB dataset, updating (inserting and deleting) 16GB data with batches of size 1GB, and executing 10,000 $ng$-approximate queries with 90\% recall@50.}
  \label{fig:ann-stream}
  \vspace{-3mm}
\end{figure}

\noindent\textbf{[Streaming scenario.]}
We finally test the performance of the indexes on streaming scenarios under workloads of different insertion/deletion/update ratios.
We skip DiskANN since the disk-version updating algorithm (i.e. FreshDiskANN~\cite{freshdiskann}) is not publicly available, and IVFPQ since it cannot reach a recall of 90\%.
For the workloads, we fix the initial dataset to be 10GB, the number of $ng$-approximate queries to be 10,000 under 90\% recall@50, and the size of the updated data to be 16GB, including insertions and deletions.
For IVF, we do not update the centroids during updating by assuming the data distribution remains the same.
In Figure~\ref{fig:ann-stream}, we show the total time of initial index building, updating and querying.
As observed, DumpyOS's overall performance is as good as the simple IVF, though DumpyOS adaptively adjusts its index structure to achieve the best point of the proximity-compactness trade-off.
With a higher deletion ratio, DumpyOS needs to make more adjustments while IVF simply marks the deleted entries.
Nevertheless, the performance difference is marginal.
\subsubsection{Discussion}
It is important to note that data series and ANN indexes are designed for different application scenarios, though the problems are conceptually the same.
ANN indexes are mainly designed for the approximate similarity search of valuable deep learning embedding vectors.
They are supposed to answer online queries in the common scenario, e.g., recommendation systems.
In this case, the fast approximate query performance is useful.

On the other hand, data series indexes are designed to manage massive data series datasets of low value-density originating from monitoring devices and natural sequences like IoT devices, etc. 
They usually serve for offline data exploration and mining tasks~\cite{ads}, such as classification and pattern recognition as we introduce in Section 1.
For example, for the large and fast IoT series data produced by sensors, a series index with \emph{high ingestion throughput} like DumpyOS is necessary~\cite{iot1,iot2,iot3}.
The same is true for data exploration tasks, where the analyst may need to build an index several times from scratch, on different data subsets.
Therefore, having solutions that achieve excellent performance in index building time and exact query answering time is very important.
Note that these abilities are \emph{not} supported by ANN indexes like DiskANN.

In a nutshell, data series and ANN indexes are the most suitable techniques of choice for their native applications (though, some technical designs can be borrowed or inspired from each other).
A more detailed discussion can be found in our previous work~\cite{bulletin-wang}.

\section{Conclusions and Future Work}
\label{sec:conclusion}
We propose a novel multi-ary data series index Dumpy with an adaptive split strategy that hits the right balance in the proximity-compactness trade-off.
By mitigating the boundary issue, Dumpy-Fuzzy and DumpyOS-F achieve even higher accuracy by checking series in the fuzzy boundary.
Moreover, the DumpyOS parallel solution fully leverages modern multi-core CPUs and NVMe SSDs, resulting in higher efficiency on both computations and I/Os.
Experiments with a variety of large, synthetic and real datasets demonstrate the efficiency, scalability, and accuracy of our solutions.



{
\begin{small}
\bibliography{journal.bib}   
\bibliographystyle{spmpsci}      
\end{small}
}
%
%

\end{document}